\documentclass{svjour3}
\usepackage{graphicx}
\usepackage{amsmath}
\begin{document}

\newcommand{\lsun}{\rm L_{\odot}}
\newcommand{\msun}{\rm M_{\odot}}
\newcommand{\rsun}{\rm R_{\odot}}
\newcommand{\me}{\dot M_{\rm Edd}}
\newcommand{\md}{\dot M_{\rm dyn}}
\newcommand{\mo}{\dot M_{\rm out}}
\newcommand{\mw}{\dot M_{\rm w}}
\newcommand{\led}{L_{\rm Edd}}
\newcommand{\rsh}{R_{\rm shock}}
\newcommand{\rinf}{R_{\rm inf}}
\newcommand{\ngc}{{NGC 4051}}
\newcommand{\pg}{{PG1211+143}}
\newcommand{\einstein}{{\it Einstein Observatory}}
\newcommand{\exosat}{{\it EXOSAT}}
\newcommand{\ginga}{{\it GINGA}}
\newcommand{\asca}{{\it ASCA}}
\newcommand{\xte}{{\it RXTE}}
\newcommand{\xmm}{{\it XMM-Newton}}
\newcommand{\chandra}{{\it Chandra}}
\newcommand{\suzaku}{{\it Suzaku}}
\newcommand{\et}{et al.\ }
\input psfig.sty

\title{Powerful Outflows and Feedback from Active Galactic Nuclei}


\author{Andrew King and Ken Pounds \\
\\
Dept of Physics and Astronomy\\
University of Leicester \\
Leicester LE1 7RH, UK}

\authorrunning{King and Pounds}
\titlerunning{AGN Outlflows and Feedback}
\maketitle

\begin{abstract}

Active Galactic Nuclei (AGN) represent the growth phases of the supermassive black holes in the center of almost every galaxy.  Powerful, highly ionized winds, with velocities $\sim
0.1- 0.2c$ are a common feature in X--ray spectra of luminous AGN, offering a plausible physical origin for the well known  connections between the hole and properties of its host.
Observability constraints suggest that the winds must be episodic, and detectable only for a few percent of their lifetimes. The most powerful wind feedback, establishing the $M
-\sigma$ relation, is probably not directly observable  at all. The $M - \sigma$  relation signals a global change in the  nature of AGN feedback. At black hole masses below $M-\sigma$
feedback is  confined to the immediate vicinity of the hole. At the $M-\sigma$ mass it becomes much more energetic and widespread, and  can drive away much of the bulge gas as a fast
molecular outflow. 
\end{abstract}


\section{INTRODUCTION}

\subsection{SMBH Scaling relations}

Astronomers now generally agree that the center of almost every galaxy but the smallest
contains a supermassive black hole (SMBH). In recent years it 
has become clear that the mass $M$ of the hole correlates
strongly with physical properties of the host galaxy. In particular the hole mass
$M$ appears always to be a fairly constant fraction of the stellar bulge mass $M_b$,
i.e.
\begin{equation}
M \sim 10^{-3}M_b
\label{mmb}
\end{equation}
(H\"aring \& Rix, 2004). Even more remarkably, observations
give a tight relation of the form
\begin{equation}
M \simeq 3\times 10^8\msun\sigma_{200}^{\alpha}
\label{msig}
\end{equation}
between the SMBH mass and the velocity dispersion $\sigma =
200\sigma_{200}~{\rm km \, s^{-1}}$ of the host galaxy's central bulge, with $\alpha \simeq 4.4\pm 0.3$
(Ferrarese \& Merritt 2000, Gebhardt et al. 2000; see Kormendy \& Ho 2013
for a recent review).
For many practical cases the relation is more conveniently written as
\begin{equation}
M \simeq 2\times 10^7\msun\sigma_{100}^\alpha
\label{msig7}
\end{equation}
with now $\sigma =100\sigma_{100}~{\rm km \, s^{-1}}$.

Since observationally determining the SMBH mass generally involves resolving its sphere of influence, of radius
\begin{equation}
R_{\rm inf} \simeq {GM\over \sigma^2} \simeq 8{M_8\over
  \sigma_{200}^2}~{\rm pc} \simeq 3{M_7\over \sigma_{100}^2}~{\rm pc},
\label{rinf}
\end{equation}
with $M = 10^8M_8\msun = 10^7M_7\msun$,
 (\ref{msig}) may represent a maximum SMBH mass for a given velocity dispersion $\sigma$ (Batcheldor 2010).
 
 \subsection{Binding Energies}

At first sight the relations (\ref{mmb},\ref{msig}) appear surprising. For (\ref{rinf}) shows that the black hole's gravity has a completely negligible effect on its host galaxy, 
which in most ways must be quite unaware of its existence. But we know (Soltan 1982) that the hole grew largely as a result of luminous accretion of gas. This released energy 
\begin{equation}
E_{\rm BH} \simeq \eta Mc^2 \sim 2\times 10^{61}M_8\,{\rm erg}, 
\label{bhbind}
\end{equation}
where $\eta \simeq 0.1$ is the
accretion efficiency, far larger than the
binding energy 
\begin{equation}
E_{\rm bulge} \sim M_b\sigma^2 \sim 8 \times
10^{58}M_8\sigma_{200}^2\,{\rm erg}
\label{bulgebind}
\end{equation}
of a host bulge of stellar mass $M_b \sim 10^3M$.
  
The vast difference in these two numbers suggests that the host must notice
the presence of the hole through its energy output, even though it is utterly insignificant in all other ways. We can already see how the black hole mass might correlate with galaxy 
properties -- the hole grows by accreting gas, but in doing this communicates some of its huge binding energy $E_{\rm BH}$ back to the gas reservoir, and so potentially limits its 
own growth. This suggests that the most relevant quantity to compare with $E_{\rm BH}$ is not 
$E_{\rm bulge}$,
but instead the gravitational binding energy of the bulge gas alone, i.e.
\begin{equation}
E_{\rm gas} = f_gE_{\rm bulge}
\label{gas}
\end{equation}
where $f_g < 1$ is the gas fraction. In the following we take this as $f_g \sim 0.16$, the cosmological mean value, giving a typical relation
\begin{equation}
E_{\rm BH} \sim 2000E_{\rm gas}
\label{BHgas}
\end{equation}
for a black hole close to the $M - \sigma$ relation (the rhs has an implicit factor $\sim \sigma_{200}^4/M_8$).

This picture requires the black hole to communicate some of its accretion energy to its host. But this process cannot be very efficient, as
otherwise the hole could disrupt the host entirely, or at the very least remove a large fraction of its gas. In this sense, the galaxy bulge leads a precarious existence. For much of 
its life it can ignore the threat that the SMBH poses, but we will see that in the end this is always decisive if accretion continues.

\subsection{Communicating the Energy: Feedback}

There are two main ways that the SMBH binding energy can potentially interact with its surroundings. By far the larger is direct radiation: after all, this is how all the accretion 
energy is initially released. But we know from observation that most light escapes relatively freely  from active galactic nuclei (AGN). This suggests that radiation is in general 
not the main way the SMBH affects its host, and we will discuss in detail why this is so in Section 7.4. The second form of coupling SMBH binding energy to a host bulge is 
mechanical. The huge SMBH accretion luminosity drives powerful gas flows into the host, making collisions and communication inevitable. One form of flow often mentioned is 
jets -- highly collimated flows
driven from the immediate vicinity of the SMBH (see Fabian 2012 for a review). To turn these into a way of affecting most
of the bulge requires a way of making the interaction relatively isotropic, perhaps with
changes of the jet direction over time. Here we will mainly consider another form
of mechanical communication which automatically has this property already. This is
the observed presence in many AGN of near--isotropic winds carrying large momentum fluxes.

\subsection{Powerful ionized winds}

Early X--ray observations of AGN yielded soft X--ray spectra frequently showing the imprint of absorption from ionized gas, the `warm absorber'; hereafter WA (Halpern 1984, Reynolds \&
Fabian 1995). More recent observations have found at least 50\% of radio-quiet AGN showing WAs in their soft X--ray ($\sim$0.3-2 keV) spectra. The limited spectral resolution of the
\einstein\ and \asca\ observations prevented  important parameters of the WAs, in particular the outflow velocity and mass rate, to be determined with useful precision. The higher
resolution and high throughput afforded by contemporary X-ray observatories, \chandra, \xmm\ and \suzaku\ has transformed that situation over the past decade, with the WA being shown,
typically, to be dominated by K-shell ions of the lighter metals (C, N, O, etc) and Fe--L, with outflow velocities of several hundred km s$^{-1}$ (Blustin \et 2005, McKernan \et
2007). 

A more dramatic discovery made possible with the new observing capabilities was the detection of blue-shifted X--ray  absorption lines in the iron K band, indicating  the presence of highly ionized
outflows with velocities $v\sim0.1-0.25c$ (Chartas \et 2002; Pounds \et 2003; Reeves \et 2003). In addition to adding an important dimension to AGN accretion studies, the mechanical
power of such winds, which for a radial flow depends on $v^{3}$, was quickly recognized to have a wider potential importance  in galaxy feedback.

Additional detections of high velocity AGN winds were delayed by the low absorption cross section of such highly ionized gas, combined with strongly blue-shifted lines in low-redshift
objects coinciding with falling telescope sensitivity above $\sim7$ keV. However, further extended observations, particularly with \xmm, found evidence in 5 additional AGN for outflow
velocities of $\sim 0.1-0.2c$ (Cappi \et 2006). Some doubts remained as to how common high velocity outflows were, as the majority of detections were of a single absorption line (with
consequent uncertainty of identification), and had moderate statistical significance, raising concerns of `publication bias' (Vaughan and Uttley 2008). In addition, only for \pg\ had a
wide angle outflow been directly measured,  confirming a high mass-rate and mechanical energy in that case (Pounds \& Reeves 2007, 2009) 

These residual doubts were finally removed following a blind search of extended AGN observations in the \xmm\ archive (Tombesi \et 2010), finding  compelling evidence in 13 (of 42)
radio quiet objects for blue-shifted iron K absorption lines, with implied outflow velocities of $\sim 0.03-0.3c$. A later search of the \suzaku\ data archive yielded a further group
of strong detections, with a median outflow velocity again $\sim 0.1c$ (Gofford et\ 2013). In addition to confirming that high velocity, highly ionized AGN winds are common, the 
yield from these archival searches shows the flows must typically have a large covering factor, and therefore be likely to involve substantial mass and energy fluxes.

The observed distributions of velocity, ionization parameter and column density are compatible with Eddington winds launched from close to the black hole, where the optical depth $\tau_{es}
\sim1$, and carrying the local escape velocity (King \& Pounds 2003). However, as the mean luminosity in most low-redshift AGN is on average sub-Eddington, such winds are likely to be
intermittent, a view supported by repeated observations and by the range of observed column densities.

For the best--quantified high-velocity outflow (the luminous Seyfert \pg), in which a wide--angle flow was directly measured (Pounds \& Reeves 2007, 2009), the wind appeared to have
more energy than needed to unbind the likely gas mass of the observed stellar bulge. This suggested that the energy coupling of wind to bulge gas must be inefficient, as seen in the
discussion following equation (\ref{BHgas}). Evidence that the fast wind in \ngc\ is shocked at a distance of $\sim 0.1$~pc from the black hole offers an explanation of why such
powerful winds do not disrupt the bulge gas: strong Compton cooling by the AGN radiation field removes
most of the wind energy  before it can be communicated. 

\section{THE OBSERVATIONAL EVIDENCE FOR ULTRA FAST OUTFLOWS}

As noted above, the requirement of X-ray observations with high sensitivity and good spectral resolution over a wide energy band delayed the discovery of powerful, highly ionized winds
from non-BAL  AGN until the launch of \chandra\ and \xmm. A decade after the first reports (Pounds \et\ 2003, Reeves \et\ 2003), high-velocity (v$\sim$0.1c), highly-ionized winds are
now established to be common in low redshift AGN.

\subsection{The fast outflow in PG1211+143}

Exploring the nature of the 'soft excess' in a sample of luminous Palomar Green AGN was a primary target in the Guaranteed Time programme awarded to Martin Turner, Project Scientist
for the EPIC Camera on \xmm\ (Turner \et 2001). At that time the X-ray spectrum in AGN above $\sim$1 keV was expected to be a rather featureless power law apart from a fluorescent
emission line at $\sim$6.4 keV from near-neutral Fe. One source, \pg, showed a surprisingly 'noisy' X-ray spectrum which one of us (KP) volunteered to explore. 

\pg, at a redshift of 0.0809 (Marziani \et\ 1996), is one of the brightest AGN at soft X-ray energies. It was classified (Kaspi et al 2000) as a Narrow Line Seyfert
1 galaxy (FWHM H$\beta$ 1800 km\, s$^{-1}$), with  black hole mass  $\sim 4\times10^{7}M_{\odot}$ and bolometric luminosity $4\times10^{45}$erg\, s$^{-1}$, indicating a mean accretion
rate close to Eddington. 

Analysis of the unusual spectral structure in the 2001 \xmm\ observation of \pg\ showed it to be  dominated by blue-shifted absorption lines of highly ionized metals, providing the
first evidence for a high velocity ionized outflow in a non-BAL AGN, with the initial identication of a deep blue-shifted Fe Lyman-$\alpha$ absorption line indicating an outflow
velocity of $\sim 0.09c$ (Pounds \et 2003). That observation, closely followed by the detection of a still higher outflow velocity from the luminous QSO PDS 456 (Reeves \et 2003),
attracted wide attention,  potentially involving the ejection of a significant fraction of the bolometric luminosity, and perhaps characteristic of AGN accreting near the Eddington
rate (King \& Pounds 2003).

Appropriately for such an unexpected discovery, the validity of the high velocity in \pg\ was not unchallenged. The near--coincidence of the observed absorption line blueshift and the
redshift of the host galaxy was a concern, notwithstanding the uncomfortably high column density of heavy metals implied by a local origin. Then, in a detailed modelling of the soft
X-ray RGS data, Kaspi \& Behar (2006) found only a much lower velocity. Any doubts relating to the absorption being local were  removed, however, by a revised velocity of $0.13-0.15c$
based on the inclusion of additional absorption lines from intermediate--mass ions (Pounds \&  Page 2006), and when repeated observations of \pg\ demonstrated that the strong Fe K absorption
line was variable over several years (Reeves \et\ 2008). 

Here we use the 2001 \xmm\ observation of \pg\ with the pn camera (Strueder \et\ 2001) to illustrate the two methods used then -- and since -- to parameterise the ionized outflow. Figure 1 shows the ratio of  EPIC
pn data to a simple power law continuum, with a deep absorption line seen near 7 keV and additional spectral structure at $\sim 1 - 4$\,keV.  

\begin{figure}                                                                                                
\centering                                                              
\includegraphics[width=8cm, angle=270]{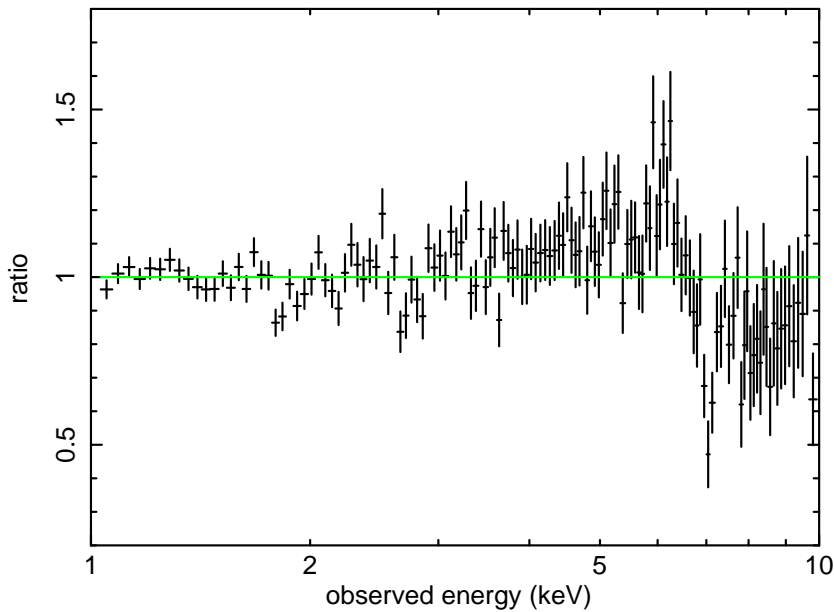}                                                                                                                                                                   
\caption                                                                
{Ratio of EPIC pn data to a simple power law continuum for the 2001 \xmm\ observation of \pg\ showing a deep absorption line near 7 keV and
additional structure between $\sim$1 and 4 keV. Deriving an outflow velocity requires the correct identification of the individual absorption lines,
which ideally requires spectral modelling with a photoionised absorber}        
\end{figure}

Fitting a negative Gaussian to the deep $\sim7$\,keV absorption line (figure 2, top panel), finds an observed line energy of $7.06\pm 0.02$\,keV, or $7.63\pm 0.02$\,keV at the AGN
redshift of 0.0809. The line is clearly resolved, with $1\sigma$ width $\sim100 \pm30$\, eV. Assuming identification with the Fe XXV resonance (6.70 keV rest energy), the blueshifted
line corresponds to an AGN outflow velocity v$\sim$0.122$\pm 0.005c$. The most likely {\it a priori} alternative identification, with the Fe XXVI Lyman--$\alpha$ line (6.97 keV rest
energy), conservatively adopted in the initial analysis (Pounds \et\ 2003), yields a lower outflow velocity $v\sim 0.095\pm 0.005c$.

An alternative procedure, which also provides additional parameters of the gas flow, requires full spectral modelling, as in Pounds \& Page (2006), and more widely in recent outflow 
studies (Section 2.3). For the 2001 \xmm\ pn spectrum of \pg, modelling the absorption from 1--10 keV with a photoionized gas derived from the XSTAR code of Kallman et al (1996)
gives an excellent fit, for a column density $N_{H}\sim 3.2 \pm 0.7\times 10^{23}$\, cm$^{-2}$, ionization parameter $\log\xi =2.7\pm 0.1\,{\rm erg\,cm\,s}^{-1}$, and an outflow
velocity (in the AGN rest frame)  $v\sim 0.149 \pm 0.003c$. The model profile (figure 2, lower panel) shows significant inner shell absorption components to the low energy wing of the 1s-2p
resonance line
which explains why simply identifying the 
absorption near 7 keV with the 6.7 keV rest energy of Fe XXV gives too low a velocity.

\begin{figure}
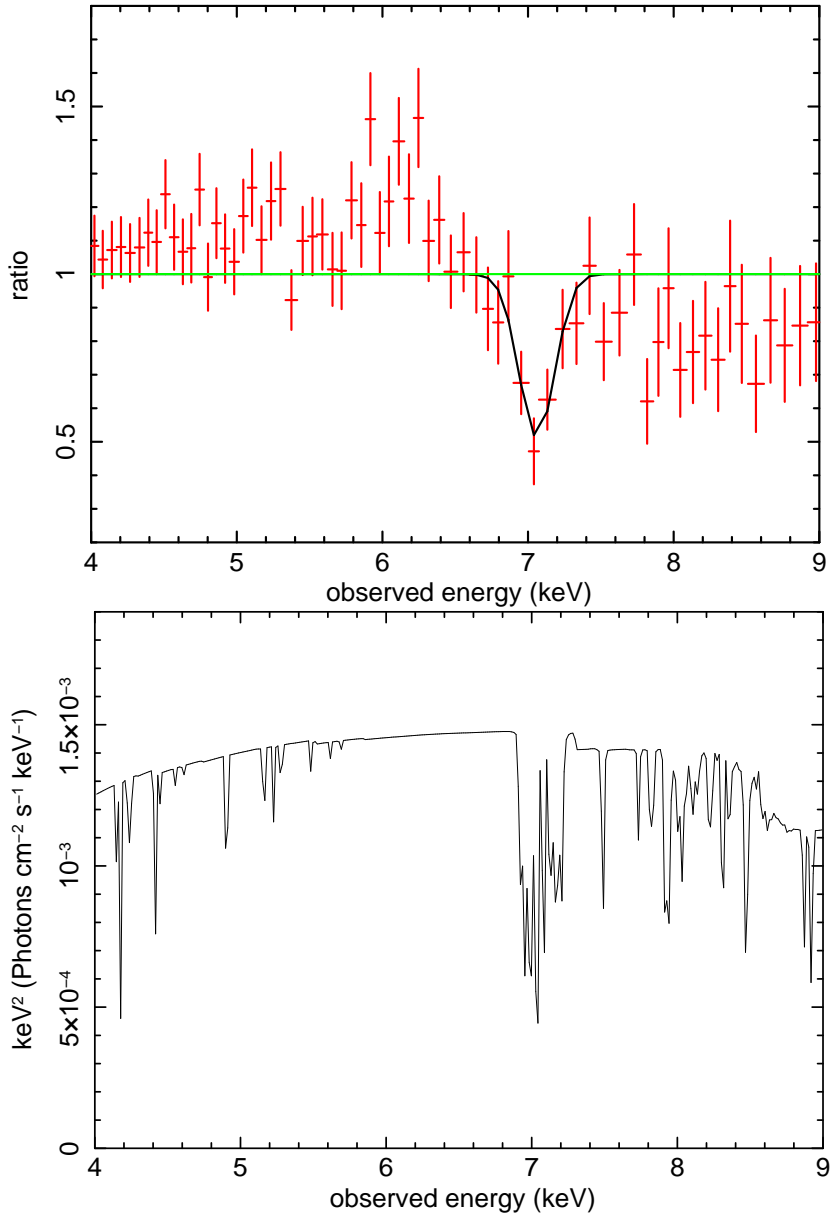
                                                                                                
\centering                                                              
\includegraphics[width=8cm, angle=270]{AR_fig2a.ps}                                                                                                                                                                   
\centering                                                              
\includegraphics[width=8cm, angle=270]{AR_fig2b.ps}                                                                                                                                                                   
\caption                                                                
{(top) A Gaussian fit to the $\sim$7 keV absorption feature finds a line energy of 7.06$\pm$0.02 keV with (1$\sigma$) width 100$\pm$30 eV.
Identification with the Fe XXV 1s-2p resonance line (6.70 keV rest energy) gives an outflow velocity $v\sim 0.12\pm 0.01c$. 
(lower) Alternative modelling with a photoionised gas over the wider 1--10 keV spectral band yields a good fit with a relatively high 
column density $N_{H}$$\sim$3.2$\pm$0.7$\times 10^{23}$ cm$^{-2}$, moderate ionisation parameter $\log\xi=2.7\pm 0.1$\, erg cm s$^{-1}$, and outflow velocity of
v$\sim$0.15$\pm$0.01c. The Fe XXV absorption line profile is seen to include lower energy components due to the addition of one or more L-shell electrons, 
showing why the simple Gaussian fit  
gives too low a velocity}        
\end{figure}

Although individually weaker than the Fe absorption, the combination of resonance lines of He-- and H--like Mg, Si, S and Ar in the broadband spectral fit is evidently driving
the spectral fit. That conclusion is confirmed with Gaussian fits to corresponding absorption features in figure 1, which find a weighted observed blueshift  of $0.055\pm 0.05$ and
outflow velocity (at the AGN redshift) of $v\sim 0.14 \pm 0.01c$, a value consistent with that found from spectral modelling, but significantly higher than from simply identifying the $\sim$7 keV absorption 
line with the resonance 1s-2p transitions of either FeXXV or FeXXVI.   

An interesting by-product of the XSTAR modelling in the above case is that the observed broadening of the $\sim7$\,keV absorption line does not require high turbulence (we used grid 25
with v$_{turb}$ of 200 km s$^{-1}$) or an accelerating/decelerating flow. Instead, intrinsically narrow absorption components remain consistent with a radial outflow, coasting
post--launch.

\subsection{Mass rate and mechanical energy in the \pg\ outflow}

Although the detection of high-speed winds in a substantial fraction of bright AGN suggests most such flows have a large covering factor, \pg\ is one of very few where a wide angle
flow has been demonstrated directly. 

Using stacked data from 4 \xmm\ observations between 2001 and  2007, Pounds \& Reeves (2007, 2009) examined the relative strength of ionized emission and absorption spectra modelled by
XSTAR to estimate the covering factor and collimation of the outflowing ionized gas. The summed pn data of \pg\ also shows a well defined P Cygni profile in the Fe K band (TBTF 3),
the classical signature of an outflow, with emission and absorption components of comparable equivalent width. Both methods indicated a covering factor  $b (=\Omega/2\pi)$ of $0.75 \pm
0.25$. Analysis of a Suzaku observation of \pg\ gives a similar result (Reeves et al. 200TBTF), with an intrinsic emission component of $\sim 6.5$ keV and width of $\sigma\sim 250$~eV, 
corresponding to a flow cone of half angle $\sim 50 \deg$, assuming velocity broadening in a radial flow. 

\begin{figure}                                                                              
\centering                                                              
\includegraphics[width=8cm, angle=270]{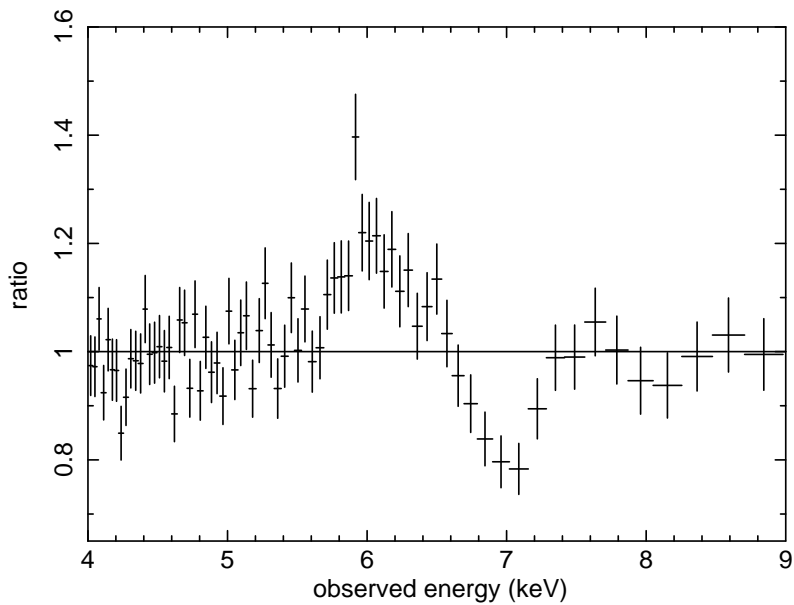} 
\caption                                                                
{The PCygni profile of Fe XXV from stacked \xmm\ pn observations of \pg\ is characteristic
of a wide angle outflow. The comparable equivalent width of emission and blue-shifted absorption components 
indicates the highly ionized outflow has a large covering factor.  From Pounds and Reeves 2009}
\end{figure}

The outflow mass rate and
mechanical energy can then be estimated, since for a uniform radial outflow of velocity v the mass rate is:

\begin{equation}
\mo \simeq {4\pi}b{nr^{2}m_p v}, 
\label{massrate}
\end{equation}
where  $n$ is the gas density at a radial
distance $r$, and $m_{p}$ is the proton mass.
 
The observed values for \pg\ find a mass loss 
rate of 
$\mo\sim 7\times10^{25}$ gm s$^{-1}$ ($ \sim 2.5\msun$~yr$^{-1}$), and mechanical energy $\sim 4.5\times 10^{44}$~erg s$^{-1}$ (Pounds \& Reeves 2009).
 
The mass loss rate is comparable to the Eddington accretion rate 
$\me = 1.3\msun$~yr$^{-1}$ for a supermassive black hole of mass $\sim 4\times 10{^7}\msun$ 
accreting at an efficiency of 10\%, while the outflow mechanical energy is only $\sim 6$\% of the Eddington luminosity, close to that predicted by continuum driving 
(equation 5 in Section 3 below).   
As noted elsewhere that energy flow rate would be more than sufficient to unbind the gas of the host galaxy bulge if all its energy were efficiently communicated.

\subsection{High speed winds are common}

The evidence for high velocity winds as an important property of AGN remained dependent on the prototype case of \pg\ for several years,  with fast outflows in two BAL AGN (Chartas \et
2002) and in the most luminous low redshift QSO PDS 456 (Reeves \et, 2003 O'Brien \et 2005) seen as rare objects. That began to change with the detection of a highly significant
outflow of velocity  $\sim 0.1$c in the Seyfert 1 galaxy IC4329A (Markowitz \et 2006), and several outflow detections in the range $\sim 0.14-0.2$c in multiple observations of Mrk 509
(Dadina \et 2005). A review in 2006 (Cappi \et 2006) listed 7 non-BAL objects with outflows of v$\sim0.1$c and several with red-shifted absorption lines.   

A major step forward came with the results of an \xmm\ archival search of bright AGN by Tombesi \et (2010), finding strong statistical evidence in 15 of 42 radio-quiet objects of
blue-shifted iron K absorption lines, identification with FeXXV or XXVI resonance absorption lines implying ultrafast outflow (UFO) velocities up to $\sim$0.3c, and clustering near
$v\sim0.1$c. A later analysis based on broad-band modelling with XSTAR photoionized grids (Tombesi \et 2011)  led to several revised velocities and confirmed that the outflows were
typically highly ionized, with $\log\xi\sim 3 - 6$\, erg cm\,s$^{-1}$,  with column densities in the range $N_{H}\sim 10^{22}-10^{24}$\,cm$^{-2}$.  A similar search of the \suzaku\
data archive (Gofford \et 2013) yielded a further group of UFO detections, finding significant absorption in the Fe K band in  20 (of 51) AGN with velocities up to $\sim 0.3c$ and a
flatter distribution than the \xmm\ sample, but a median value again v$\sim0.1$c.

\begin{figure*}                                                                                                
\centering                                                              
\includegraphics[width=7.5cm, angle=0]{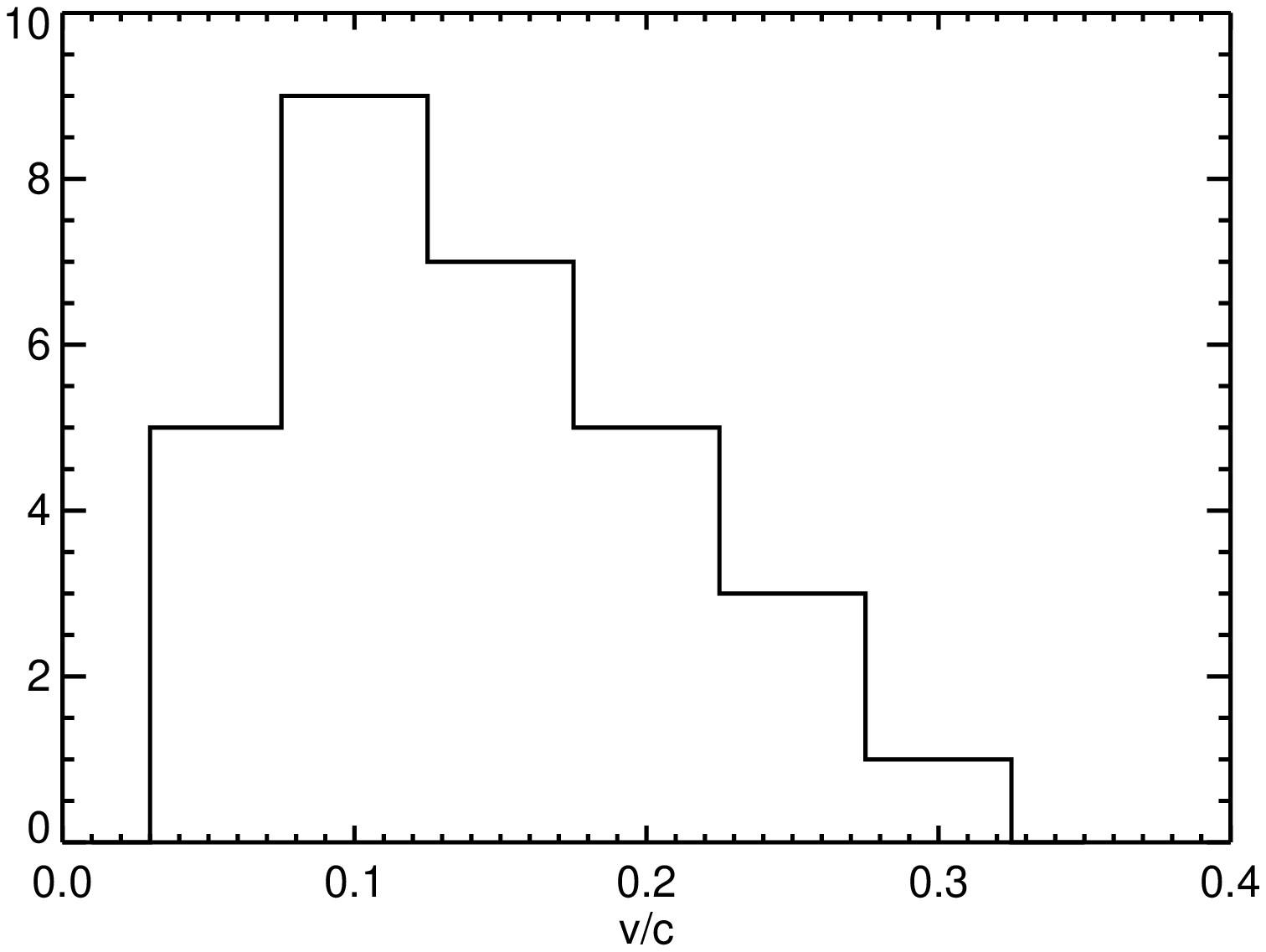}                                                                                                                                                        
\begin{center}
\hbox{
 \hspace{0.1 cm}
   \includegraphics[width=6 cm]{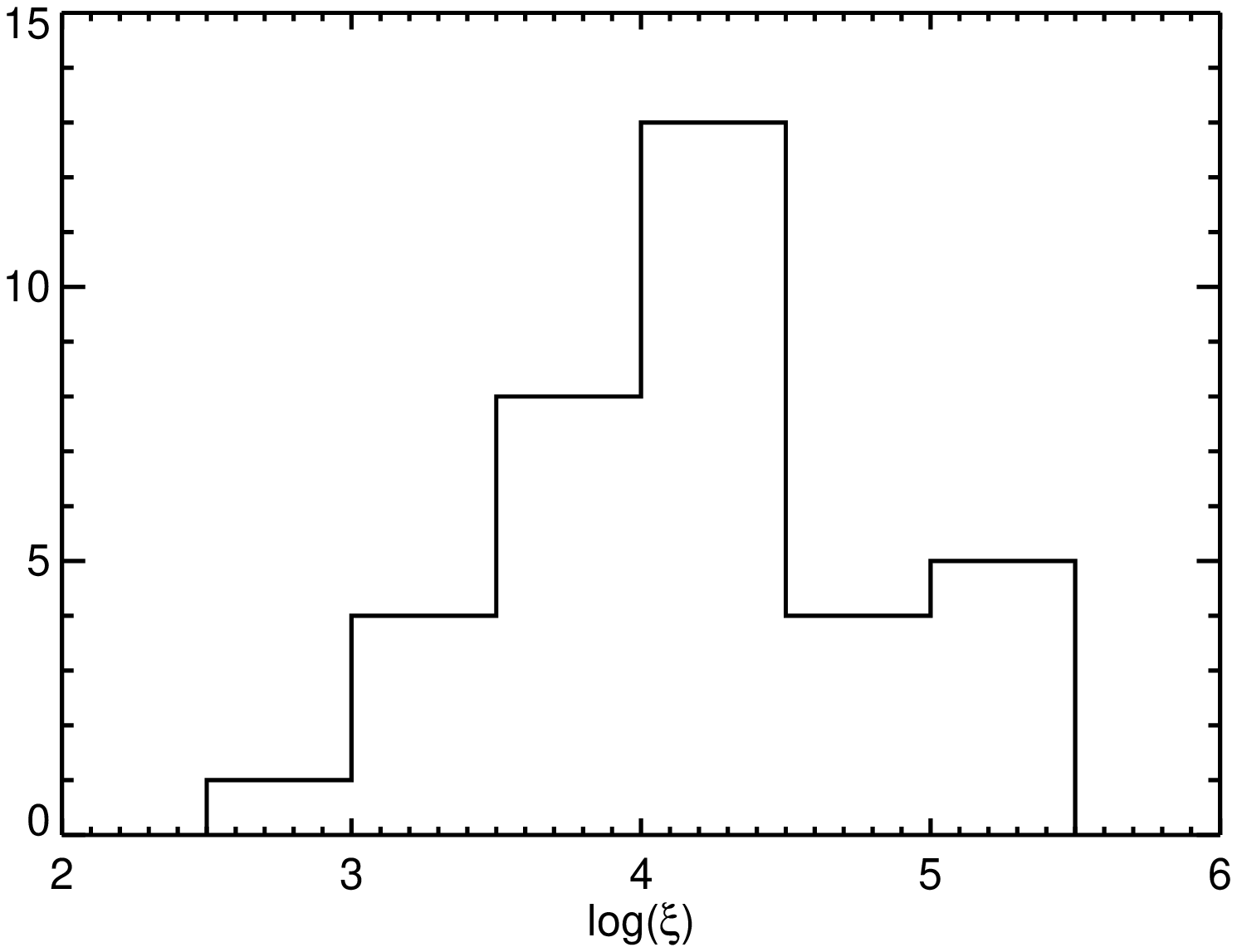}
 \hspace{0.1 cm}
 \includegraphics[width=6 cm]{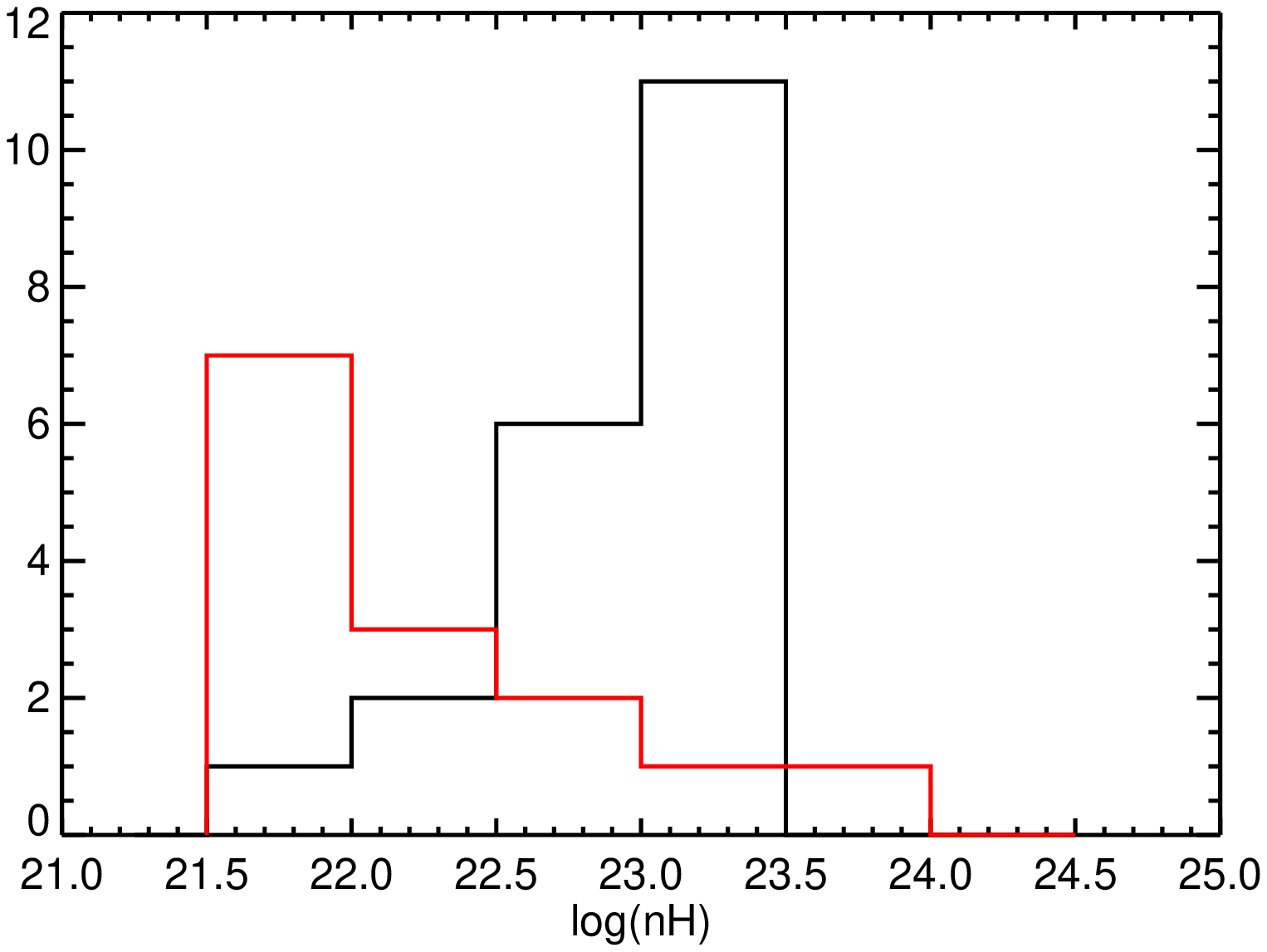}} 
   \end{center} 
\caption
{Distribution of outflow velocities, ionization parameter (erg\,cm\, s$^{-1}$) and column density (cm$^{-2}$) obtained from modelling the individual spectra from 
extended observations of type 1 AGN in the \xmm\ and \suzaku\ data
archives (Tombesi \et 2011, Gofford \et\ 2013). 
The red lined histogram refers to lower limits in column density} 
\end{figure*}

Figure 4 brings together the results from the spectral modelling analyses of the \xmm\ and \suzaku\ surveys. We follow Tombesi \et\ (2011) in defining UFOs as having outflow velocities
greater than $10^{4}$~km\, s$^{-1}$, to discriminate against WAs or post-shock flows (Sections 2.4 and 4). The velocity plot shows a peak at $\sim 0.1$c, with a tail extending to
$\sim0.3$c.     In terms of the continuum-driving Black Hole Winds model (King \& Pounds 2003) the higher velocities would imply a higher value of the accretion efficiency $\eta$,
with the future potential for such observations to  provide a measure of black hole spin. Equation (22) also suggests the low velocity tail in both the Tombesi \et\ and Gofford \et\
distributions could relate to primary outflows formed at a higher accretion ratio (but see Section 3.1).

Figure 4 also shows the distribution of ionization parameter and absorption column density from the surveys of Tombesi \et\ (2011) and Gofford \et\ (2013). The high ionization parameter,
peaking near $\log\xi\sim 4$, explains why the detection of UFOs has been almost exclusively from X--ray observations in the Fe K band, leaving open the possibility that fully ionized
outflows (also consistent with continuum driving) will become detectable when the AGN luminosity (and hence ionization) falls. In assessing observational data it is important to note that for a radial outflow the
observed column density is a line--of--sight integration over the flow time, dominated by the higher density at small radii, while the ionization parameter is governed by the current
AGN luminosity. The column density, which generally lies below $N_{H}\sim 10^{24}$~cm$^{-2}$, can vary rapidly and turns out to be a powerful diagnostic of the flow history and
dynamics. We return to the observability of UFOs in Section 3.3.

\subsection{Evidence for a shocked flow}

The mechanical energy in a fast wind, such as that in \pg, was noted in Section 2.2 to be incompatible with the continued growth of the black hole and stellar bulge of the host galaxy,
unless the flow is short-lived or the coupling of wind energy to bulge gas is highly inefficient. A recent \xmm\ observation of the narrow-line Seyfert galaxy \ngc\ has provided the first evidence of a fast ionized wind
being shocked, with subsequent strong cooling leading to most of the initial flow energy being lost before it can be communicated to the bulge gas. We outline a possible scenario for that event below.

\ngc\ was found in the \xmm\ archival search to have a high velocity wind during an observation in 2002 when the source was in an unusually low state, the initial identification with
Fe XXVI Lyman-$\alpha$ in Tombesi \et (2010) indicating a velocity of $\sim$0.15c. In a full spectral fit (Tombesi \et 2011) identification with Fe XXV was preferred, with an increased
velocity $\sim$0.20c. Significantly, in a 2001 observation of \ngc,  when the X-ray flux was much higher, a strong outflow was detected at $\sim$6000 km s$^{-1}$, but no ultra-fast
wind was seen.

\begin{figure}                                                                                                
\centering                                                              
\includegraphics[width=8cm, angle=270]{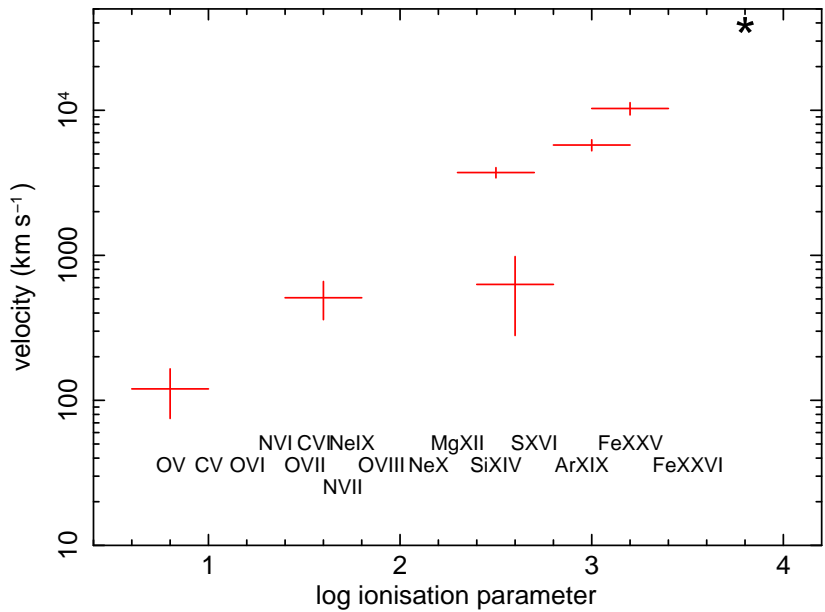}                                                                                  
\caption                                                                
{The outflow velocity and ionization parameters for 6 XSTAR photoionised absorbers used to fit the RGS and EPIC
spectra of \ngc,  together with a high point representative of the pre-shock wind, show the linear correlation expected for a
mass-conserved cooling flow (see Pounds and King 2013)}        
\end{figure}

It seems that the detection of a UFO in \ngc\ is unusually dependent on the source flux, with evidence for a high velocity wind ($v\sim 0.13c$) again found only during periods when the
ionizing continuum was low during a further \xmm\ observation in 2009 (Pounds \& Vaughan 2012). An additional factor may be the low redshift (z=0.00234)  of \ngc, which makes a high
velocity wind more difficult to detect with current observing facilities.

The 600 ks \xmm\ observation of the Seyfert 1 galaxy \ngc\ in 2009, extending over 6 weeks and 15 spacecraft orbits, broke new ground by finding an unusually rich absorption spectrum
with  multiple outflow velocities, in both RGS (den Herder \et\ 2001) and EPIC spectra, up to $\sim 9000~{\rm km\,s}^{-1}$ (Pounds \& Vaughan 2011a). Inter-orbit variability is seen in both absorption and
emission lines, with strong recombination continua (RRC) and velocity-broadened resonance lines providing constraints on the dynamics and geometry of the putative post-shock flow (Pounds and
Vaughan 2011b, 2012).

\subsection{A self-consistent model for the shocked wind in \ngc}

More complete modelling of both RGS and EPIC pn absorption spectra of \ngc\  found a highly significant correlation of outflow velocity and ionization state (figure 5), as expected
from mass conservation in a post--shock flow (King 2010, Pounds \& King 2013). The additional analysis also found a range of column densities to be required by the individual XSTAR 
absorption components, suggesting an inhomogeneous shocked flow, perhaps with lower ionization gas clumps or filaments embedded in a more extended, lower density and more highly
ionized flow. 

Theoretical considerations suggested a key factor in determining the structure of the post-shock flow was likely to be the cooling time, as discussed in more detail in Section 4.  In
particular, the fate of a fast wind depends on the distance it travels before colliding with the ISM or slower-moving ejecta, with Compton cooling dominating for a shock occurring
sufficiently close to the AGN continuum source.

\begin{figure*}
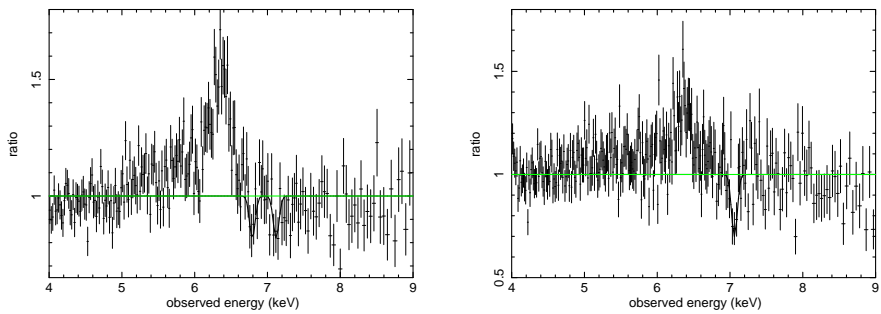

\begin{center}
\hbox{
 \hspace{0.1 cm}
   \includegraphics[width=4cm, angle=270]{AR_fig6l.ps}
 \hspace{0.15 cm}
   \includegraphics[width=4cm, angle=270]{AR_fig6r.ps}} 
   \end{center} 
\caption
{Fe K profiles from observations of \ngc\ several days apart show an increased level of ionization coinciding with a hard X-ray flare (data from Pounds and Vaughan 2012). The ratio of 
resonance absorption lines of Fe XXV and Fe XXVI is a sensitive measure of the ionization state of the absorbing gas} 
\end{figure*}

Importantly, flux-linked variations in the ratio of FeXXV to Fe XXVI absorption in the 2009 \xmm\ observation (figure 6) provided a measure of the Compton cooling time, the mean flow
speed  then determining the shell thickness of the hotter, more highly ionized  flow component. The detection of strong recombination continua (RRC) in the soft X-ray spectra
furthermore suggested an increasing density in the decelerating post-shock flow, with two-body cooling becoming increasingly important.  

To pursue that idea we note that at the (adiabatic) shock the free--free (thermal bremsstrahlung) and Compton cooling times are
\begin{equation}
t_{\rm ff} \simeq 3\times 10^{11}{T^{1/2}\over N}~{\rm s}  = 20{R_{16}^2\over M_7\dot m}~{\rm yr}
\label{ff}
\end{equation}
and
\begin{equation}
t_{\rm C} = 10^{-4}{R_{16}^2\over M_8}~{\rm yr}
\label{compt}
\end{equation}
respectively (see King et al. 2011: here $T, N$ are the postshock temperature and number density, $R_{16}$ is the shock radius 
in units of $10^{16}$~cm,
$M_7$ is the black hole mass in units of $10^7M_{\odot}$, 
and $\dot m \sim 1$ is the Eddington factor of the mass 
outflow rate). 

After the adiabatic shock, the gas cools rapidly via inverse Compton cooling, while its density rises as $N \propto T^{-1}$ (pressure is almost constant in an isothermal shock), and 
\begin{equation}
t_{\rm ff}\propto {T^{1/2}\over N} \propto T^{3/2}, 
\end{equation}
which means that the free--free cooling time decreases sharply while the Compton time does not change. Eventually free--free (and other 
atomic two--body processes) become faster than Compton when $T$ has decreased sufficiently below the original shock temperature $T_s \sim 1.6\times10^{10}$~K. 
From (\ref{ff}, \ref{compt}) above this requires
\begin{equation}
\left({T\over T_s}\right)^{3/2} < 5\times 10^{-5}
\end{equation}
or 
\begin{equation}
T < 2\times 10^7~{\rm K}.
\end{equation}

The temperature of ionization species forming around a few keV is therefore likely to be determined by atomic cooling processes rather than Compton cooling. The strong recombination
continua in \ngc\ (Pounds \& Vaughan 2011b, Pounds \& King 2013) are direct evidence for that additional cooling, with the RRC flux yielding an emission measure  for the related flow
component. In particular, the onset of strong two-body cooling results in the lower-ionization, lower-velocity gas being confined in a relatively narrow region in the later stages
of the post-shock flow. The structure and scale of both high and low ionization flow regions can be deduced from the observations and modelling parameters. 

For the highly ionized post-shock flow, the iron Ly--$\alpha$ to He--$\alpha$ ratio will be governed by the ionizing continuum and recombination time.  Significant variations in this
ratio are found on inter-orbit timescales (Pounds \& Vaughan 2012), with an example shown in figure 6. For a mean temperature of $\sim$1 keV, and recombination  coefficient of
$4.6\times10^{-12}$~cm$^3$\,s$^{-1}$  (Verner \& Ferland 1996), the observed recombination timescale of $\sim 2\times10^{5}$ s corresponds to an average particle density of $\sim
4\times10^{6}$\,cm$^{-3}$. Comparison with a relevant absorption column $N_{H}\sim 4\times10^{22}$~cm$^{-2}$ from the XSTAR modelling indicates a column  length scale of $\sim 10^{16}$
cm.  Assuming a mean velocity of the  highly ionized post-shock flow of 6000 km s$^{-1}$, the observed absorption length corresponds to a flow time  $\sim 1.7\times10^{7}$ s (0.6 yr).
Equation (11) finds   a comparable cooling time for \ngc\ at a shock radius $R \sim 10^{17}$~cm.

\begin{figure*}                                                                                                
\centering                                                              
\includegraphics[width=11cm]{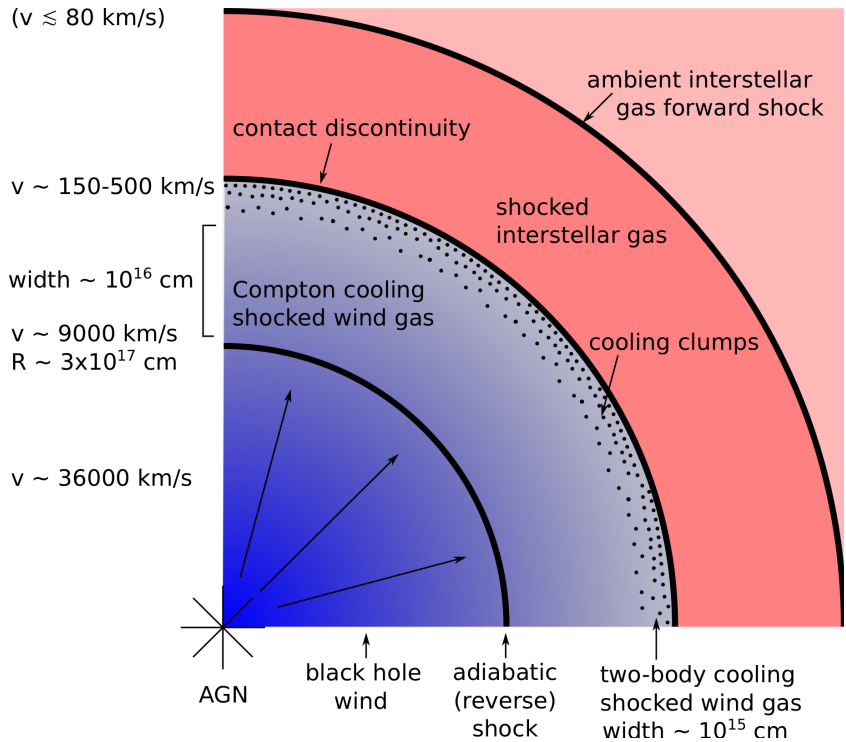}                                                                                  
\caption                                                                
{Schematic view of the shock pattern resulting from the impact of a black hole wind (blue) on the interstellar gas (red) of the host galaxy. The accreting  supermassive black hole drives a
  fast wind (velocity $v \sim \eta c/\dot m \sim 0.1c$), whose ionization state makes it observable in X--ray absorption lines. It collides with the ambient gas in the host galaxy and is
  slowed in a strong shock. The inverse Compton effect from the quasar's radiation field rapidly cools the shocked gas, removing its thermal energy and strongly compressing and slowing it
  over a very narrow radial extent. In the most compressed gas, two--body cooling becomes important, and  the flow rapidly cools and slows over an even narrower region. In \ngc\ this region
  is detected in the soft X--ray  spectrum, where absorption (and emission) are dominated by the lighter metals. The cooled gas exerts the preshock ram pressure on the galaxy's interstellar
  gas and sweeps it up into a dense shell (`snowplow'). The shell's motion then drives a milder outward shock into the ambient interstellar medium. This shock ultimately stalls unless the
  SMBH mass has reached the value $M{_\sigma}$ satisfying the $M - \sigma$ relation (from Pounds and King 2013)} 
 \end{figure*} 

For the low-ionization flow component, decay of strong RRC of NVII, CVI and CV (Pounds \& Vaughan 2011b,  Pounds \& King 2013), occurs over $\sim
2-6$ days. With an electron temperature from the mean RRC profile of $\sim 5$~eV, and recombination  coefficient for CVI of $\sim 10^{-11}$~cm$^3$\,s$^{-1}$  (Verner \& Ferland 1996),
the observed RRC decay timescale corresponds to a (minimum) electron  density of $\sim$$2\times10^{6}$~cm$^{-3}$.   A column density of $1.5\times10^{21}$~cm$^{-2}$ from modelling
absorption in the main low-ionization flow component then  corresponds to an absorbing path length of $7\times10^{14}$~cm.  

The RRC emission flux provides a consistency check on the above scaling.  Assuming solar abundances, and 30 percent of recombinations direct to the ground state, a CVI RRC
flux of $\sim 10^{-5}$ photons~cm$^{-2}$ s$^{-1}$  corresponds to  an emission measure of $\sim 2\times10^{62}$~cm$^{-3}$, assuming a Tully--Fisher distance to \ngc\ of 15.2 Mpc. 
With a mean particle density of $\sim2\times10^{6}\,{\rm cm}^{-3}$ the emission volume ($4\pi R^{2}\Delta R$) is  $\sim 5\times10^{49}\,{\rm cm}^{3}$ . Assuming a spherical shell
geometry of the flow, with fractional solid angle $b$, shell thickness $\Delta R\sim 7\times10^{14}\,{\rm cm}$, and shell radius $R\sim 10^{17}$\,cm, the measured RRC flux is reproduced
for $b\sim 0.5$. 

Although this excellent agreement may be fortuitous given the approximate nature and averaging of several observed and modelled parameters, the mutual consistency  of
absorption and  emission of the photoionized flow is encouraging. Given that only blueshifted RRC emission is seen, $b\sim$0.5 is consistent with a wide-angle flow, visible only on the
near side of the accretion disc.

Figure 7 illustrates the main features of the overall \ngc\ outflow, a fast primary wind being shocked at a radial distance of order 0.1pc, within the zone of influence of an SMBH of
$1.7\times 10^{6}\msun$. The initially hot gas then cools in the strong radiation field of the AGN, with a Compton cooling length determining the absorption columns of Fe and the other
heavy metal ions. Two--body recombination provides additional cooling as the density rises downstream, eventually becoming dominant. Absorption (and emission) in the soft X--ray band
is located primarily in this thinner outer layer of the post--shock flow. 

It is interesting to note that similar shocking of fast outflows provides a natural link between UFOs and the equally common `warm absorbers' in AGN (Tombesi \et 2013). While the onset
of strong two--body cooling, resulting in the intermediate column densities being small, might explain why evidence for intermediate-flow velocities has awaited an unusually long
observation of a low mass AGN, the accumulated `debris' of shocked wind and ISM could be a major component of the `warm absorber'. See Section 7.4.1 for a discussion.

\subsection{Variability of UFOs}

While it is likely that powerful winds blow continuously in AGN in rapid growth phases, it is important to note that the existing observations of UFOs are restricted to bright, low-redshift 
AGN, $z\leq 0.1$, where the X-ray fluxes are sufficient to yield high quality spectra. Repeated observation of several bright AGN frequently show changes in the equivalent
width in the primary Fe K absorption line. 

Variability in the strength of blueshifted Fe-K absorption over several years in \pg\ was first noted in a comparison of the initial \xmm\ and \chandra\  observations (Reeves \et
2008), and confirmed by repeated \xmm\ observations (Pounds and Reeves 2009). Multiple observations of the luminous Seyfert 1 galaxy Mrk 509 (Cappi \et 2009) found variations in both intensity and blueshift 
of Fe K absorption lines. The archival searches provide the most comprehensive variability data, with repeated observations of several AGN demonstrating that
variability of absorption line equivalent width (EW) over several years is common. More rapid variability in EW, over a few months, is reported in the \xmm\ archive for Mrk 509, Mrk 79 and
Mrk 841, with both velocity and EW change in $\leq$2 days for Mrk766. 

In addition, the `hit rate' of UFOs for multiply-observed AGN in the archival \xmm\ data search (Tombesi \et 2010) was relatively low, being 1 of 6 observations for NGC4151, MCG-6-30-15 (0/5), Mrk509 (3/5),
\ngc (1/2), Mrk79 (1/3), Mrk205 (1/3), and Mrk290 (1/4). Overall, though 101 suitably extended observations yielded 36 narrow absorption line detections in the Fe K band, only 22 were
observed at $> 7$~keV. While the UFO `hit rate' of $\sim$ 22\% is a lower limit set by the sensitivity of available exposures it seems clear that the fast outflows
currently being detected in low-redshift AGN are far from continuous.

\section{BLACK HOLE WINDS}

\subsection{The Eddington Accretion Ratio in AGN}

We have seen from Section 2 that a large fraction of observed AGN show ultrafast outflows. Unless we view every AGN from a very particular angle (so implying a much larger total population) this must mean that these winds have large solid angles $4\pi b$ with $b\sim 1$, i.e. they are quasispherical. We recall that UFOs are observed to have total scalar momenta $\sim L_{\rm Edd}/c$, where $\led$ is the Eddington luminosity of the SMBH. We can argue that on quite general grounds, SMBH mass growth is likely to occur at accretion rates close to the value $\me = \led/\eta c^2$ which would produce this luminosity. As we noted earlier, the Soltan (1982) relation shows that the largest SMBH gained most of their mass by luminous
accretion, i.e. during AGN phases. But the fraction of AGN among all
galaxies is small, strongly suggesting that when SMBH grow, they
are likely to do so as fast as possible. The maximum possible rate of accretion from 
a galaxy bulge with velocity dispersion $\sigma$
is the dynamical value
\begin{equation}
\dot M_{\rm dyn} \simeq {f_g \sigma^3\over G},
  \label{dyn}
\end{equation}
where $f_g$ is the gas fraction. This rate applies when gas which was previously in gravitational equilibrium is disturbed and falls freely, since one can estimate that the gas mass
was roughly $M_g \sim \sigma^2f_gR/G$. Once this is destabilized it must fall inwards on a dynamical timescale $t_{\rm dyn} \sim R/\sigma$. This gives the result (\ref{dyn}), since $\dot M_{\rm dyn} \sim M_g/t_{\rm dyn}$.

Numerically we have
\begin{equation}
\md \simeq 280 \sigma_{200}^3~\msun\, {\rm yr}^{-1}
\label{dyn2}
\end{equation}
where we have taken
$f_g = 0.16$, the cosmological baryon fraction of all matter. 
We have
\begin{equation}
\me = {L_{\rm Edd}\over \eta c^2} = {4\pi GM\over \kappa\eta c}
\label{edd}
\end{equation}
where $L_{\rm Edd}$ is the Eddington luminosity and $\kappa$ is the
electron scattering opacity. With $\eta = 0.1$ and
black hole masses $M$ close to the observed $M - \sigma$
relation (\ref{msig}) we find
\begin{equation}
\me \simeq 4.4~\sigma_{200}^4\msun\, {\rm yr}^{-1}
\end{equation}
and an Eddington accretion ratio 
\begin{equation}
\dot m < {\md \over \me} \simeq {64\over \sigma_{200}} \simeq
     {54\over M_8^{1/4}}.
\label{eddrat}
\end{equation}

Thus even dynamical infall cannot produce extremely super--Eddington accretion rates on to supermassive black holes.  But the rate (\ref{dyn}) is already a generous overestimate, since it
assumes that the infalling gas instantly loses all its angular
momentum. Keeping even a tiny fraction of this instead forces the gas to orbit the black hole and form an accretion disc, which slows things down drastically. Gas spirals
inwards through a disc on the viscous timescale
\begin{equation}
t_{\rm visc} = {1\over \alpha}\left({R\over H}\right)^2\left({R^3\over
  GM}\right)^{1/2}
\label{tvisc1}
\end{equation}
where $\alpha \sim 0.1$ is the Shakura--Sunyaev viscosity parameter,
while the disc aspect ratio $H/R$ is almost constant with radius, and typically
close to $10^{-3}$ for an AGN accretion disc (e.g. Collin--Souffrin \&
Dumont 1990). Then $t_{\rm visc}$ approaches a Hubble time
even for disc radii of only $1$~pc. Gas must evidently be rather accurately channelled towards the SMBH in order to accrete at all, constituting a major problem for theories of how AGN are fuelled. 

Given all this, we see that while it is possible for AGN accretion rates to reach Eddington ratios 
$\dot m \sim 1$, significantly larger ones are unlikely unless the SMBH mass is far below the 
$M - \sigma$ value appropriate to the galaxy bulge it inhabits. In other words, only
relatively modest values $\dot m \sim 1$ of the Eddington ratio are likely in SMBH growth episodes. 

Indirect evidence supporting this view comes from stellar--mass compact binary systems. The dynamical rate is relatively much larger here, as the equivalent of  (\ref{dyn}) is $\dot M \sim v_{\rm orb}^3/G \sim
M_2/P$, with $v_{\rm orb}$ the orbital velocity of a companion star in a binary of period $P$. This implies rates approaching a solar mass per few hours in many cases, if dynamical accretion ever occurs. 
These systems have highly super--Eddington apparent luminosities, probably as the result of geometric collimation (cf King
et al. 2001).  But significantly there are no obvious AGN analogues of the ultra-luminous X-ray sources (ULXs), suggesting that Eddington ratios $\dot m \gg 1$ are very unusual or absent in AGN.

\subsection{Eddington Winds}

Given this, we can crudely model the UFOs discussed in Section 2 as
quasi-spherical winds from SMBH accreting at modest Eddington ratios $\dot m = \dot
M/\me \sim 1$.  Winds like this have electron
scattering optical depth $\tau \sim 1$, measured inwards from infinity
to a distance of order the Schwarzschild radius
(cf eq. \ref{nh2} below). So on average every
photon emitted by the AGN scatters about once before escaping to infinity.  Since electron scattering is front--back symmetric, each photon on average gives up all its momentum to the wind, and so
the total (scalar) wind momentum should be of order the photon
momentum, or
\begin{equation}
\mw v \simeq {L_{\rm Edd}\over c},
\label{mom}
\end{equation}
where $v$ is the wind's terminal velocity. The winds of hot stars obey relations like this.
For accretion from a disc, as here, the classic paper of Shakura \&
Sunyaev (1973) finds a similar result at super--Eddington mass inflow rates: the excess accretion is
expelled from the disc in a quasispherical wind.

Equation (\ref{edd}) now directly gives the wind terminal velocity as
\begin{equation}
v \simeq {\eta\over \dot m}c \sim 0.1c.
\label{v}
\end{equation}
From eq. (\ref{v}) we get the instantaneous wind mechanical luminosity as
\begin{equation}
L_{\rm BH\,wind} =\mw {v^2\over 2} \simeq {\led\over c}{v\over 2} \simeq {\eta\over
2}\led
\simeq 0.05\led.
\label{lum}
\end{equation}
This relation turns out to be highly significant (see Sections 5.3, 7.3). 

Ohsuga \& Mineshige (2011) show in detail that winds with these properties
(their Models A and B) are a natural outcome of mildly super--Eddington accretion. In
particular their Model A and B winds are predicted (cf their Figure 3) to have mechanical luminosities $\sim 0.1\led$, in rough agreement with equation (\ref{lum}). 

Compared with the original disparity $E_{\rm BH} = \eta Mc^2 \sim 2000 E_{\rm gas}$ between black hole and bulge gas binding energies outlined in the Introduction,  we now have a relation 
\begin{equation}
E_{\rm BH\,wind} \simeq {\eta^2\over 2}Mc^2 \sim 100E_{\rm gas} 
\label{windbulge}
\end{equation} 
between the available black hole wind mechanical energy and the bulge binding energy. Although the mismatch is less severe, it still strongly suggests that the bulge gas would be massively disrupted if 
it experienced the full mechanical luminosity emitted by the black hole for a significant time. So the coupling of mechanical energy to the host ISM cannot be efficient all the time (see the discussion 
in Section 2). We show how this works in Section 4.

\subsection{Observability}

As we have seen, observations frequently give the hydrogen column density $N_H$ through a UFO wind from the X--ray absorption spectrum. We can show that this quantity determines whether a given UFO wind is observable or not. Using (\ref{mom})  in the mass conservation equation
\begin{equation}
\mw = 4\pi br^2v\rho(r),
\label{mass}
\end{equation}
where $\rho(r)$ is the mass density, we find the equivalent hydrogen column density of the wind as
\begin{equation}
N_H \simeq \int_{R_{\rm in}}^{\infty} {\rho\over m_p}{\rm d}r = \int_{R_{\rm in}}^{\infty}{\mw\over 4\pi r^2 bv}{\rm d}r = {\led\over 4\pi bm_pR_{\rm in}cv^2}\  ,
\label{nh}
\end{equation}
where $R_{\rm in}$ is its inner radius, $m_p$ is the proton mass,  and we have used (\ref{mom}) at the last step. From the definition of $\led$ we find the wind electron scattering optical depth 
\begin{equation}
\tau =  N_H \sigma_T\simeq {GM\over bv^2R_{\rm in}}
\label{nh2}
\end{equation}
with $\sigma_T \simeq \kappa m_p$ the Thomson cross--section.
This shows self--consistently that the
scattering optical depth $\tau$ of a continuous wind is $\sim 1$ (cf King \& Pounds
2003, equation 4) at the launch radius $R_{\rm launch} \simeq GM/bv^2 = (c^2/2bv^2)R_s \simeq 50R_s$. 

The measured values of $N_H$ (Tombesi et al. 2011, Gofford \et 2013, Figure 4) are always smaller than the value $N_H \simeq 1/\sigma_T \simeq 10^{24}~{\rm cm}^{-2}$ for a continuous wind, 
and actually lie in the range  $N_{22} \sim 0.3 - 30$, where $N_{22} = N_H/10^{22}~{\rm cm}^{-2}$.  It is perhaps not surprising that observations do not show any UFO systems with 
$N_H > 10^{24}~{\rm cm}^{-2}$. These AGN would be obscured at all photon energies by electron scattering, and perhaps difficult to see. Although such systems might be common, we probably cannot detect them. To have a good chance of seeing  a UFO system we need a smaller $N_H$, so from (\ref{nh2}) the inner surface $R_{\rm in}$ of the wind must be larger than $R_{\rm launch}$. This is only possible if all observed UFOs are episodic, i.e. we see them some time after the wind from the SMBH has switched off. In this sense UFOs are more like a series of sporadically--launched quasispherical shells than a continuous outflow. The $N_H$ value of each shell is dominated by the gas near its inner edge (cf (\ref{nh}), so we probably at most detect only the inner edge of the most recently--launched shell. We can quantify this by setting $R_{\rm in} = vt_{\rm off}$, where $t_{\rm off}$ is the time since the launching of the most recent wind episode ended. Using (\ref{nh2}) gives
\begin{equation}
t_{\rm off} = {GM\over bv^3 N_H\sigma_T} \simeq {3M_7\over bv_{0.1}^3N_{22}}~{\rm months},
\label{toff}
\end{equation}
where $v_{0.1} = v/0.1c$.

As seen in Figure 4, all UFOs have $N_{22} \sim 0.3 - 30$, and most of the SMBH masses are $\sim 10^7\msun$. Evidently the launches of most observed UFO winds halted weeks or months before
the observation. At first sight this is surprising. The strength of the characteristic blueshifted absorption features defining UFOs is closely related to $N_H$. These features would be still stronger if there
were UFOs with $N_{22} > 100$, but none are seen. We note from (\ref{toff}) that observing a UFO like this would require us to catch it within days of launch. Given the relatively sparse coverage of X--ray
observations of AGN this is unlikely. So the apparent upper limit to the observed $N_H$ may simply reflect a lack of observational coverage, and implies that most UFOs are short-lived 

The lower limit to $N_H$ in the Tombesi et al. sample is also interesting. Once $N_H$ is smaller than some critical value, any blueshifted absorption lines must become too weak to detect. The strongest are the
resonance lines of hydrogen-- and helium--like iron, which have absorption cross--sections $\sigma_{\rm Fe} \simeq 10^{-18}$~cm$^2$. Given the abundance by number of iron as $Z_{\rm Fe} = 4\times 10^{-5}$ times
that of hydrogen, the condition that one of these lines should have significant optical depth translates to $Z_{\rm Fe}N_H\sigma_{\rm Fe} > 1$ or $N_{22} > 2.5$. This is similar to the lowest observed values.
From (\ref{toff}) this means that current observations cannot detect UFO winds launched more than a few months in the past, because the blueshifted iron lines will be too weak. Even these observed UFOs should gradually
decrease their $N_H$ and become unobservable if followed for a few years. We see in Section 5 that the UFO wind typically travels $\sim 10M_7$~pc or more before colliding with the host galaxy's interstellar
gas, which takes $t_{\rm coll}/v \sim 300M_7v_{0.1}^{-1}$~yr. Finally, a UFO may be unobservable simply because it is too strongly ionized, so that no significant $N_H$ can be detected.

All this means that the state of the AGN seen in a UFO detection does not necessarily give a good idea of the conditions required to launch it. In particular, the AGN may be observed at a sub--Eddington
luminosity, even though one might expect luminosities $\sim \led$ to be needed for launching the UFO. This may be the reason why AGN showing other signs of super--Eddington phenomena (e.g. narrow--line Seyfert 2
galaxies) are nevertheless seen to have sub--Eddington luminosities most of the time (e.g. NGC 4051; Denney et al. 2009): the rather short wind episodes are launched in very brief phases in which accretion is
slightly super--Eddington, whereas the long--term average rate of mass gain may be significantly sub--Eddington.

In summary, it is likely that current UFO coverage is remarkably sparse. We cannot see a continuous wind at all. We can only see an episodic wind shell shortly after launch, and then only for a tiny fraction $t_{\rm off}/t_{\rm coll}
\sim 10^{-3}v_{0.1}^{-2}N_{22}^{-1}$ of its  $\sim 300 - 3000$~yr journey to collision with the host ISM. So it seems that the vast majority of UFO wind episodes remain undetected: more AGN must produce them
than we observe, and the known UFO sources may have far more episodes than we detect. 

All this has important consequences for how we interpret observations in discussing feedback. The most serious is that the most powerful form of feedback -- from AGN at the Eddington limit producing continuous
winds --  is probably not directly observable at all.

\subsection{The Wind Ionization State, and BAL QSOs}

The ionization parameter 
\begin{equation} 
\xi = {L_i\over NR^2}
\label{ion}
\end{equation}
essentially fixes the ionization state of a black hole wind wind, and so determines which spectral lines are observed. Here $L_i = l_iL_{\rm Edd}$ is the ionizing luminosity, with
$l_i< 1$ a dimensionless parameter specified by the quasar spectrum,
and $N = \rho/\mu m_p$ is the number density of the UFO gas.  We use (\ref{v}, \ref{mass}) to get
\begin{equation}
\xi = 3\times 10^4\eta_{0.1}^2l_2\dot m^{-2} = 3\times 10^4v_{0.1}^2l_2
\label{ion2}
\end{equation}
where $l_2 = l_i/10^{-2}$, and $\eta_{0.1} = \eta/0.1$. 

This relation shows how the wind momentum and mass rates determine its ionization parameter and so its line spectrum as well as its speed $v$. Given a quasar spectrum $L_{\nu}$ , the ionization state has to
arrange that the threshold photon energy defining $L_i$, and the corresponding ionization parameter $\xi$, together satisfy (\ref{ion2}). This shows that the excitation must be high: a low threshold photon
energy (say in the infrared) would imply a large value of $l_2$, but then (\ref{ion2}) gives a high value of $\xi$ and so predicts the presence of very highly ionized species, physically incompatible with such
low excitation.

For suitably chosen continuum spectra (\ref{ion2}) has a range of solutions. A given spectrum might in principle allow more than one solution, the applicable one being specified by initial conditions. For a
typical quasar spectrum, an obvious self--consistent solution of (\ref{ion2}) is $l_2 \simeq 1$, $\dot m \simeq 1$, $\xi \simeq 3\times 10^4$. Here  the quasar radiates the Eddington luminosity. We can also
consider situations where the quasar's luminosity has decreased after an Eddington episode but the wind is still flowing, with $\dot m \simeq 1$. Then the ionizing luminosity $10^{-2}l_2L_{\rm Edd}$ in
(\ref{ion2}) is smaller, implying a lower $\xi$.  As an example, an AGN of luminosity $0.3L_{\rm Edd}$ would have $\xi \sim 10^4$.  This gives a photon energy threshold appropriate to FeXXV and Fe XXVI 
(i.e. $h\nu_{\rm threshold}\sim 9$~keV). We conclude that Eddington winds from AGN are likely to have velocities $\sim 0.1c$, and show the presence of helium-- or hydrogen-like iron in accord with the absorption
reported in Section 2. Zubovas \&
King (2013) show that this probably holds even for AGN which are significantly sub--Eddington.


We can see from (\ref{v}) that a larger Eddington factor $\dot m$ is likely to
produce a slower wind. From comparison with ULXs (see Section 3.1) we also expect the AGN radiation to be beamed away from a large fraction of the UFO, which should therefore be be less ionized, and as a result more easily detectable than the small fraction receiving the beamed radiation. These properties -- slower, less ionized winds -- characterize BAL QSO
outflows, perhaps suggesting that
systems with larger $\dot m > 1$ appear as BAL QSOs.  Zubovas \& King (2013)
tentatively confirm this idea. 

\section{THE WIND SHOCK}

\subsection{Momentum-- and Energy--Driven Flows}

So far we have only studied the black hole wind. But we know that  this wind must have a significant effect on the host galaxy when it impacts directly on its interstellar medium (ISM).  In this Section we model
the wind and host ISM as roughly spherically symmetric, and consider the effects of deviations from this simple picture later. 

The pattern of the wind--ISM interaction (Figure 7) is qualitatively identical to that of a stellar wind hitting the interstellar medium around it (see e.g. Dyson \& Williams 1997).  The black hole wind (shown
in blue) is abruptly slowed in an inner (reverse) shock where the temperature approaches $\sim 10^{11}$~K if ions and electrons reach equipartition (but see the discussion below). The shocked wind gas acts like
a piston, sweeping up the host ISM  (shown in red) at a  contact discontinuity moving ahead of it. Because this swept--up gas moves supersonically into the ambient ISM, it drives an outer (forward) shock into it
(see Figs. 7 and 8 [top]).

\begin{figure*}
  \centering
    \includegraphics[width= 11cm]{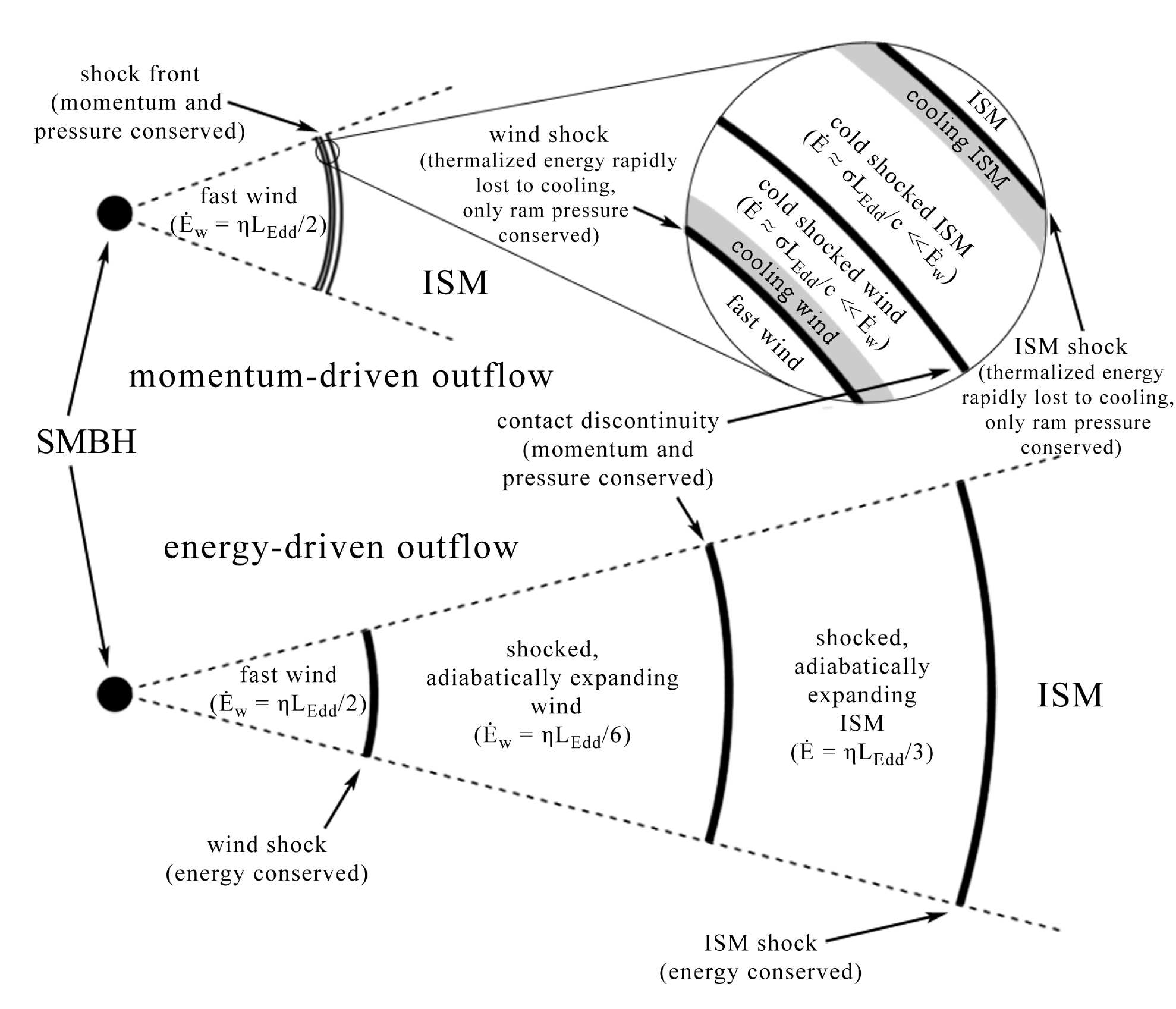}
  \caption{Schematic picture of momentum--driven (top) and
    energy--driven (bottom) outflows. In both cases a fast wind
    (velocity $\sim 0.1c$) impacts the interstellar gas of the host
    galaxy, producing an inner reverse shock slowing the wind, and an
    outer forward shock accelerating the swept--up gas. In the
    momentum--driven case (top), corresponding to the UFOs discussed in Section 2, 
    the shocks are very narrow and rapidly cool
    to become effectively isothermal. Only the ram pressure is
    communicated to the outflow, leading to very low kinetic energy
    $\sim (\sigma/c)\led$. 
    In an energy--driven outflow (bottom), the shocked
    regions are much wider and do not cool. They expand adiabatically,
    communicating most of the kinetic energy of the wind to the
    outflow (in simple cases divided in a ratio of about 1:2 between
    the shocked wind and the swept--up gas). The outflow radial
    momentum flux is therefore greater than that of the
    wind. Momentum--driven conditions hold for shocks confined to
    within $\sim 1$~kpc of the AGN, and establish the $M - \sigma$
    relation (\ref{msig2}) (King, 2003; King, 2005). Once the
    supermassive black hole mass attains the critical $M - \sigma$
    value, the shocks move further from the AGN and the outflow
    becomes energy--driven. This produces the observed large--scale
    molecular outflows which probably sweep the galaxy clear of
    gas. (From Zubovas \& King, 2012a).}
  \label{fig:outflow}
\end{figure*}

The dominant interaction here is the reverse shock slowing the black hole wind, and injecting energy into the host ISM.  The nature of this shock differs sharply depending on whether or not some form of cooling
(typically radiation) removes significant energy from the hot shocked gas on a timescale shorter than its flow time. If the cooling is strong in this sense (`momentum--driven flow'), most of the preshock kinetic
energy is lost (usually to radiation). The very rapid cooling means that the shocked wind gas is highly compressed, making the postshock region geometrically narrow (see the upper part of Figure 8). This kind of
narrow, strongly cooling region ia often idealised as a discontinuity, know as an `isothermal shock' (cf Dyson \& Williams, 1997). As momentum must be conserved, the postshock gas transmits just its ram pressure
(\ref{mom}) to the host ISM. We will see that this amounts to transfer of only a fraction $\sim \sigma/c \sim 10^{-3}$ of the mechanical luminosity $E_{\rm BH\,wind} \simeq 0.05\led$ (cf equation \ref{lum}) to
the ISM. In other words, in the momentum--driven limit, only energy 
 
\begin{equation} 
E_{\rm mom} \sim {\sigma\over c}E_{\rm BH\,wind} \sim {\sigma\over c}{\eta^2\over 2}Mc^2 \sim 5\times 10^{-5}Mc^2 \sim
0.1E_{\rm gas} 
\label{momen}  
\end{equation} 
is injected into the bulge ISM, i.e. about 10\% of the bulge gas binding energy $f_gM_b\sigma^2$ for black holes close to the $M - \sigma$ relation (there is now an
implicit factor $\sigma_{200}^5/M_8$ on the rhs). Thus momentum--driven flows do not threaten the bulge's integrity. Indeed we will see that they never interact with most of it, so there is no danger that the
black hole will drive away the gas and suppress accretion. A momentum--driven regime is a stable environment for black--hole mass growth.

In the opposite limit where cooling is negligible, the postshock gas retains all the mechanical luminosity   \begin{equation} E_{\rm wind} \simeq 0.05\led \simeq 100 E_{\rm gas} \label{energy} \end{equation} (cf
equation \ref{windbulge}) thermalized in the shock, and instead expands adiabatically into the ISM. The postshock gas is now geometrically extended (see the lower part of Figure 8), unlike the momentum--driven
(`isothermal') case. This `energy--driven flow' is much more violent than momentum--driven flow. The estimate (\ref{energy}) is for a black--hole mass near the $M - \sigma$ relation; a hole with  mass a factor
of 100 below this would already unbind the bulge in doubling its mass. Unless the shock interaction is markedly aspherical, a black hole in an energy--driven environment is unlikely to reach observed SMBH
masses.

Given these starkly different outcomes we must decide under what conditions we have momentum-- or energy--driven outflows. Simple estimates immediately show that ordinary atomic two--body processes have no significant effect in cooling the wind shock. But the wind shock is exposed to the radiation field of an Eddington--accreting supermassive black hole. This has a characteristic temperature of no more than $\sim 10^7$~K, far lower than the
wind's immediate post-shock temperature of $\sim 10^{10} - 10^{11}$~K. Electrons in the postshock gas lose energy to these photons through the inverse Compton effect (cf Ciotti \&
Ostriker, 1997), at a rate dependent on the radiation density.
For wind shocks close to the SMBH, the accretion radiation field is
intense enough that this effect cools the postshock wind gas in less
than the momentum--driven flow time $\sim R/\sigma$ (see below), and we are
self--consistently in the momentum--driven regime, provided that the postshock gas is in equipartition, i.e. that electron and ion temperatures remain effectively equal. (We consider this further in Section 4.2 below.)
 
For shocks at larger radii $R$ the radiation energy density decreases as
$R^{-2}$, increasing the cooling time as $R^2$. The flow time increases only as $R$, so for $R$ greater than the critical cooling radius

\begin{equation}
R_C \sim 500M_8^{1/2}\sigma_{200}~{\rm pc}
\label{rcool}
\end{equation}

(King, 2003, 2005; King et al 2011, Zubovas \& King 2012b) the cooling time is longer than the flow time, and the flow must be energy--driven.  
So we have the general result that momentum--driven flows are confined to a small region $R < R_C$, while energy--driven flows must be large--scale. This is just as one would expect, given that a momentum--driven flow allows stable black hole mass growth, while an energy--driven one is likely to expel most of the bulge gas.

It is plausible then that the observed UFO winds can lead to
momentum--driving through strongly cooled shocks close to the
SMBH. In this picture all but a small fraction of the mechanical luminosity
(\ref{lum}) of the black hole wind is eventually radiated away as an inverse Compton continuum
with characteristic photon energy $\sim 1$~keV. Pounds \& Vaughan
(2011) report a possible detection of this spectral component in
the Seyfert 1 galaxy NGC 4051. 
As required for consistency, the luminosity of this component is comparable to the
expected mechanical luminosity (\ref{lum}) of the wind in that system. 

Cooling shocks are called `isothermal' because the
gas temperature rapidly returns to something like its preshock value. Momentum conservation requires that the gas is also strongly slowed and compressed as it cools. So the postshock velocity of the X--ray
emitting gas should correlate with its temperature (or roughly,
ionization) while Compton cooling is dominant. Once this has
compressed the gas sufficiently, two--body processes such as
free--free and bound--free emission must begin to dominate, since they
go as the square of the density, and their cooling times decrease with
temperature also (Pounds \& King 2013). Section 2.5 above shows
that there is direct observational evidence for both of these
effects in NGC 4051. So this object (uniquely) shows three signatures of a cooling shock: an inverse Compton continuum, an ionization--velocity correlation, and the appearance of two--body processes in the spectrum.

\subsection{Shock Cooling}

Cooling (or the lack of it) has a defining effect on the physics of the interaction between the black hole wind and the host ISM, so we must check the simple picture above.  In particular 
the inverse Compton effect acts only on electrons, but the energy of the postshock gas is initially almost all in its ions. We assumed above that the electron and ion temperatures quickly come into equipartition after the shock, allowing the inverse Compton effect to drain energy from the ions. 

This assumption can be questioned. Faucher--Gigu\`ere \& Quataert (2012) show that if the only process coupling electrons and ions is Coulomb collisions, there is a significant parameter space where
equipartition does not occur.  
although they do not rule out substantial momentum--conserving phases. An important
consideration here is that many processes other than direct Coulomb collisions may rapidly equilibrate electron and ion temperatures. Faucher--Gigu\`ere \& Quataert (2012) attempt to put limits on the incidence
of such collisionless coupling by appealing to observations of the solar wind, but this is an area of considerable physical uncertainty. 

Another way of using observations to decide if shock cooling is effective is to look for the inverse Compton spectral component directly revealing the cooling. Bourne \et\ 2013) argue that the apparent lack 
of such a component in most AGN spectra rules out cooling shocks. 
But we recall from Section 3.3 that the coverage of UFOs is extremely sparse.
Actually observing a collision 
and so the inverse Compton emission
is inevitably a very rare event. 
It appears that observationally ruling out Compton shock cooling is so far inconclusive.

\section{THE $M - \sigma$ RELATION}

\subsection{Reaching $M - \sigma$: the Momentum--Driven Phase}

We are now equipped to discuss the impact of a UFO on the host interstellar gas. 
We already noted that the wind impact implies a pair of shocks each side of the
contact discontinuity between the wind and the host ISM. 
Initially the wind shock is close the hole, and we assume that inverse 
Compton cooling from the AGN radiation field cools it rapidly and
puts the flow in the momentum--driven regime. The 
region of gas between the wind shock and
the contact discontinuity, where it impacts and sweeps up the host ISM,
is very narrow (cf the upper panel of Figure 8). The outer shock accelerating the ISM is also strongly cooled, so that the `snowplow' region of swept--up ISM is narrow as well. So we can 
treat the whole region between the inner and outer shocks as a single narrow,
outward--moving gas shell, whose mass grows as it sweeps up the host ISM (see Fig. 9).

As a simple model of a bulge, we assume that the matter 
of the host galaxy is distributed with an isothermal profile
of velocity dispersion $\sigma$, with mass density
\begin{equation}
\rho(r) = {f_g\sigma^2\over 2\pi Gr^2},
\label{rho}
\end{equation}
so that the mass within
radius $R$ is
\begin{equation}
M(R) = {2\sigma^2 R\over G}.
\label{iso}
\end{equation}
A distribution like this is expected if the bulge results from mergers.
For a roughly constant gas fraction $f_g$, the mass of the narrow
swept--up gas shell at radius $R$ is 
\begin{equation}
M_g(R) = {2f_g\sigma^2 R\over G}
\label{gas}
\end{equation}
so that the shell has the equation of motion
\begin{equation}
{{\rm d}\over {\rm d}t}[M_g(R)\dot R] + {GM_g(R)[M + M(R)]\over R^2} =
{\led\over c},
\label{motion1}
\end{equation}
where $M$ is the SMBH mass.
From (\ref{iso}, \ref{gas}) and the definition of $\led$ (equation \ref{edd})
this simplifies to
\begin{equation}
{{\rm d}\over {\rm d}t}(R\dot R) + {GM\over R} = -2\sigma^2\left(1 -
{M\over M_{\sigma}}\right),
\label{motion2}
\end{equation}
where
\begin{equation}
M_{\sigma} = {f_g\kappa\over\pi G^2}\sigma^4.
\label{sig}
\end{equation}
Multiplying through by $R\dot R$ and integrating once gives the first integral
\begin{equation}
R^2\dot R^2 = -2GMR - 2\sigma^2\left[1 - {M\over M_{\sigma}}\right]R^2
+ \   {\rm constant}
\label{int}
\end{equation}
For large $R$ we have
\begin{equation}
\dot R^2 \simeq  - 2\sigma^2\left[1 - {M\over M_{\sigma}}\right]
\label{vel}
\end{equation}
which has no solution for $M < M_{\sigma}$. Physically this says that if the SMBH mass is
below $M_{\sigma}$ the swept--up shell of interstellar gas cannot reach large radius because the
Eddington thrust of the black hole wind is too small lift its weight
against the galaxy bulge potential. 
The SMBH cannot remove the gas from its surroundings, and goes on accreting.
Any gas shell it drives outwards eventually becomes too
massive, and so tends to fall back and probably fragment. This is likely to stimulate star formation in the shell remnants.

The precise value $M_{\sigma}$ depends on the average gas fraction
$f_g$. For a protogalaxy forming at high redshift
we expect $f_g = \Omega_{\rm
  baryon}/\Omega_{\rm matter} \simeq 0.16$ (Spergel et al.,
2003). Galaxies forming at later times may have larger $f_g$ 
if they have gained a lot of gas, or smaller $f_g$ if they have been largely swept clear of gas, or have turned a lot of their gas into stars. 
With the gas fraction $f_g$ fixed at
the cosmological value $f_c = 0.16$, the expression
\begin{equation}
M_{\sigma} = {f_g\kappa\over \pi G^2}\sigma^4 \simeq 3.2\times
10^8\msun\sigma_{200}^4
\label{msig2}
\end{equation}
is remarkably close to the observed relation (\ref{msig}), even though
it contains no free parameter.  We shall see in Section 5.5 why observations tend to give an exponent of $\sigma$ slightly larger than the value 4 derived here. This agreement strongly suggests that
SMBH growth stops at this point, although we must do some more work
to show this (see Sections 5.2, 5.3 and 5.5 below).

The derivation here took the simplest possible description of a
galaxy spheroid as an isothermal sphere (cf equation \ref{iso}). We should ask
if things change significantly if the galaxy bulge is more complicated than this. If the potential is spherically symmetric but
the cumulative mass $M(R)$ is not simply linear in $R$, we still get a first integral of the equation of motion (\ref{motion1}) simply by multiplying through by $M(R)\dot R$, giving the giving the condition for a swept--up momentum--driven shell to reach large radii. Relations very like (\ref{msig2}) emerge in each case, so we expect qualitatively similar behavior. McQuillin \& McLaughlin (2012) show this explicitly for three widely--used density distributions (Hernquist 1990; Navarro et al. 1996,
1997; and Dehnen \& McLaughlin 2005):
the results
are in practice scarcely distinguishable from (\ref{msig2}). 

Whatever the bulge geometry, the black hole always communicates its presence only through the ram pressure of its wind, so we are always dealing with strongly radial forces in the solid angles exposed to this wind (this is not true of gas pressure, as we shall see in Section 5.3). It is likely that the orientation of the accretion disc with respect to the host galaxy changes with each new episode of accretion (so--called chaotic accretion, King \& Pringle 2006, King et al. 2008), tending to isotropize the long--term effect of momentum feedback.
Together with the sudden huge increase in the spatial scale as the critical black hole mass is reached (see the next Section) this may explain why the simple spherically symmetric prediction (\ref{msig2}) seems to give a surprisingly accurate estimate of the critical mass.

\subsection{What Happens When $M = M_{\sigma}$?}

The result (\ref{msig2}) is so close to observations that it strongly suggests that feedback somehow cuts off the growth of the black hole at a mass very close to this value. 
Some feeding may continue from gas in the immediate vicinity of the hole which is too dense to be affected by the ram pressure of the black hole wind. A thin accretion disc has this property for example, but cannot have a gas mass larger than $\sim (H/R)M \ll M$ without fragmenting and forming stars. 

But we still have to explain
precisely how the gas is expelled. For example, one might worry that although momentum--driving can push the ISM away and inhibit central accretion, some kind of infall and SMBH accretion might restart shortly after momentum--driving is switched off, perhaps leading to alternating stages of quiescence and growth, eventually to masses far above $M_{\sigma}$. Observations show that black hole accretion occurs preferentially in gas--rich galaxies (cf Vito et al. 2014), so it seems
that the black hole must largely clear the galaxy bulge of gas to terminate its growth.
We will see later (Section 7) that if no other process than momentum--driving operated, this requirement would indeed lead to black hole masses significantly larger than $M_{\sigma}$, in conflict with observation. 

This last point means that the way black hole growth influences the host galaxy must change radically when $M = M_{\sigma}$. It is straightforward to see why it should.
We saw above that for $M < M_{\sigma}$ the Eddington thrust cannot push the wind shocks to large $R$. As a result the wind shock remains efficiently Compton cooled, enforcing momentum--driving. It follows that the SMBH cannot stifle its own growth if $M < M_\sigma$. But all this changes once the SMBH exceeds the critical mass (\ref{msig2}). Now even for a very small increment ($O(R_{\rm inf}/R_C)
\sim 10^{-2})$ of $M$ above $M_{\sigma}$, a momentum--driven shell can reach the critical
radius $R_C$. Crucially, this means that the wind shocks are no longer efficiently cooled: they become energy--driven. The shocked wind gas can now use all of its energy to push the interstellar gas as it expands into the host bulge. This motion becomes explosive and rapidly reaches kiloparsec lengthscales, comparable with the size of the bulge itself, rather the much smaller (parsec) scales of the momentum--driven phase. 

So the real significance of the $M - \sigma$ relation is that it marks the point where outflows undergo a global transition from momentum-- to energy--driving.

\subsection{Clearing Out a Galaxy: the Energy--Driven Phase}

We know from (\ref{windbulge}) that an energy--driven outflow
has more than enough energy to remove the interstellar gas entirely, and so presumably suppress further SMBH growth. Here we examine how this works in detail.

Once $M > M_{\sigma}$ the outflow geometry changes completely (see
Figure 8). The shocked wind region is no longer narrow (as in the upper panel of Figure 8),
but large and expanding because of its strong thermal pressure (lower panel of Figure 8). 
The shocked wind's thermal expansion pushes its shock inwards where it must hover at the  cooling radius $R_C$ (Zubovas et al. 2013).  If it tries to move within $R_C$, momentum driving instantly pushes it out again (remember $M > M_{\sigma}$). 

The shocked wind rapidly evens out its internal pressure as it expands at its sound speed  $\sim 0.03c$, so we take this pressure $P$ as uniform over this region (but changing with time).  
The contact discontinuity at the outer edge of the shocked wind sweeps up the surrounding shocked ISM as before, but now has the equation of motion
\begin{equation}
{{\rm d}\over {\rm d}t}\biggl[M_g(R)\dot R\biggr] + {GM_g(R)M(R)\over R^2} = 
4\pi R^2P,
\label{motion2}
\end{equation}
where the pressure $P$ is much larger than the ram pressure $\rho v^2$ appearing in (\ref{motion1}). In the second term on the lhs we have neglected the contribution $GM_gM/R^2$ of the black hole gravity, as $R \gg R_C > R_{\rm inf}$.
To make the problem determinate we need the energy equation. 
(This did not appear explicitly in the momentum--driven case because it was equivalent to the defining condition that all the wind energy not associated with the ram pressure was rapidly lost to radiation.)
Here the energy equation constrains the pressure
$P$ by specifying the rate that energy is fed into the shocked gas,
minus the rate of $P{\rm d}V$ working on the ambient gas and against
gravity:
\begin{equation}
{{\rm d}\over {\rm d}t}\biggl[{4\pi R^3\over 3}.{3\over 2}P\biggr] = 
{\eta\over 2}\led - P{{\rm d}\over {\rm d}t}\biggl[{4\pi\over
  3}R^3\biggr]
- 4f_g{\sigma^4\over G}.
\label{Eenergy}
\end{equation}
We take a specific heat ratio $\gamma = 5/3$, use
(\ref{lum}) for the energy input from the outflow and (\ref{iso})
to simplify the gravity term $GM(R)M(R)/R^2$.  
Now we use (\ref{motion2}) to eliminate $P$ from (\ref{Eenergy}), and
replace the gravity terms as before using the isothermal expression for $M(R)$. We take the AGN luminosity as $l\led$ to allow for small deviations from the Eddington value. This gives
\begin{equation}
{\eta\over 2}l\led = \dot R{{\rm d}\over {\rm d}t}\biggl[M(R)\dot R\biggr]
+ 8f_g{\sigma^4\over G}\dot R + 
{{\rm d}\over {\rm d}t}\biggl\{{R\over 2}\dot R{{\rm d}\over {\rm d}t}
\biggl[M(R)\dot R\biggr] + 2f_g{\sigma^4\over G}R\biggr\}
\end{equation}
and so
\begin{equation}
{\eta\over 2}l\led = 
{2f_g\sigma^2\over G}\biggl\{ {1\over 2}R^2\dddot{R} + 3R\dot R\ddot R
+ {3\over 2}\dot R^3\biggr\} + 10f_g{\sigma^4\over G}\dot R.
\label{motion4}
\end{equation}
This describes the motion of the interface (`contact discontinuity' in the lower panel of Figure 8) between wind and interstellar
gas in the energy--driven case, replacing equation (\ref{motion1}) in the momentum--driven case.

The energy--driven regime applies as soon as the SMBH mass reaches $M_{\sigma}$, and we will see that the host ISM is now quickly removed. 
We assume $M= M_{\sigma}$ in $\led$, and see that (\ref{motion4}) has a  solution $R = v_e t$ with
\begin{equation}
2 \eta lc = 3{v_e^3\over \sigma^2} + 10v_e
\end{equation}
The assumption $v_e << \sigma$ leads to a contradiction ($v_e \simeq
0.01c >> \sigma$), so
\begin{equation}
v_e \simeq \biggl[{2\eta l\sigma^2c\over 3}\biggr]^{1/3} \simeq
925l^{1/3}\sigma_{200}^{2/3}~{\rm km\ s}^{-1}
\label{ve}
\end{equation}

This solution is an attractor. Figure 9 shows that all solutions quickly converge to it, regardless of initial conditions.
Physically, its meaning is that if shock cooling is ineffective, the extra gas
pressure accelerates the previously momentum--driven gas shell to
this new higher velocity. Figure 9 also confirms that if the 
driving by the AGN switches off when the contact discontinuity is at radius $R_0$, it decelerates as predicted by the
analytic solution of (\ref{motion4}) with $\led = 0$ found by King et
al. (2011):
\begin{equation}
\dot R^2 = 3\biggl(v_e^2 + {10\over 3}\sigma^2\biggr)\biggl({1\over
  x^2} - {2\over 3x^3}\biggr) - {10\over 3}\sigma^2
\label{dotr}
\end{equation}
where $x = R/R_0\geq 1$.
Noting that $v_e$ depends only weakly (as  $v_e \sim l^{1/3}$)
on the luminosity, these results show that fluctuations -- or even the intermittent disappearance -- of the AGN luminosity have almost no effect on the outflow once it has started, because the flow still  has a large reservoir of thermal energy available for driving. In particular an outflow can persist long after the central AGN has turned off, and the real agency driving an observed outflow may have been an AGN even if this is currently observed to be weak or entirely absent.


\begin{figure}
  \centering
    \includegraphics[width=9cm]{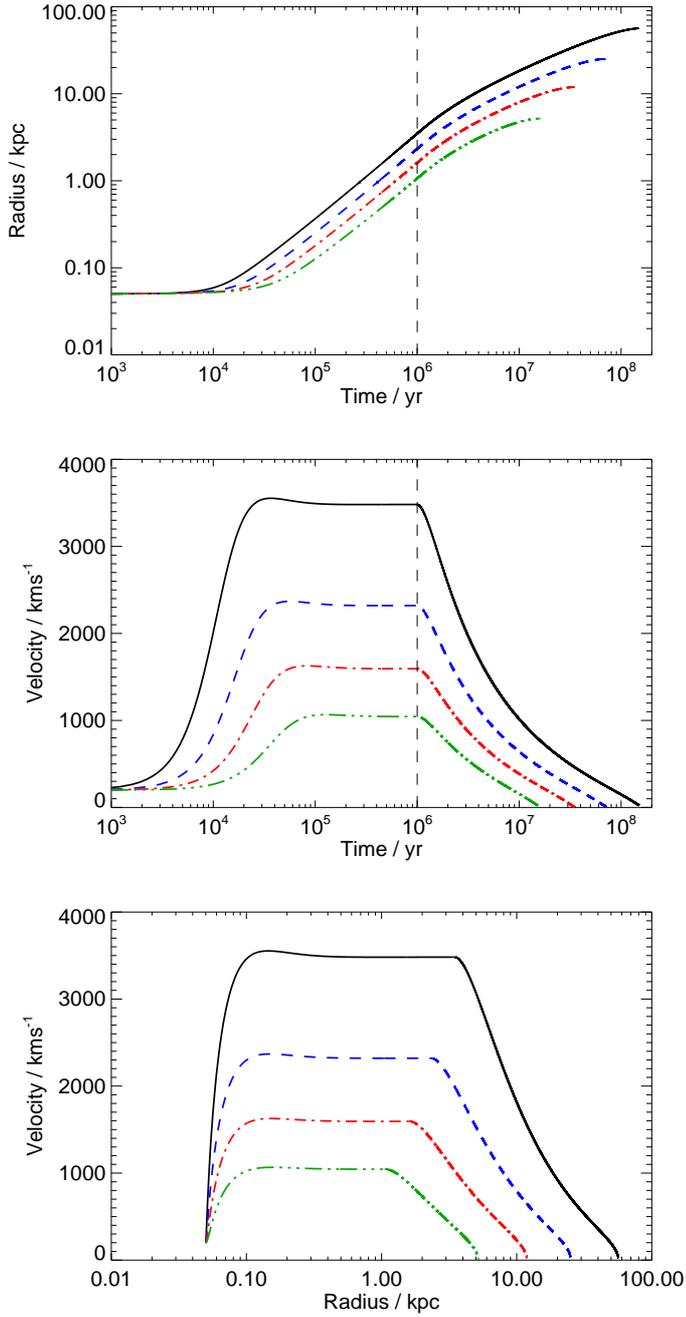}
    \caption{Evolution of an energy--driven shock pattern for the case
      $\sigma = 200$~km\,s$^{-1}, f_g = 10^{-2}$ computed numerically
      from the full equation (\ref{motion4}). Top: radius vs time,
      middle: velocity vs time, bottom: velocity vs radius. The curves
      refer to different initial conditions: black solid -- $R_0 = 10$
      pc, $v_0 = 400$~km\,s$^{-1}$; blue dashed -- $R_0 = 100$ pc, $v_0 =
      1000$~km\,s$^{-1}$;
      red dot--dashed -- $R_0 = 50$ pc, $v_0 = 200$~km\,s$^{-1}$.
      All these solutions converge to the attractor
      (\ref{ve}). The vertical dashed line marks the time $t = 10^6$
      yr when (for this case) the quasar driving is switched off.  All
      solutions then follow the analytic solution (\ref{dotr}). A case
      where the quasar remains on for a Salpeter times $\sim 4 \times
      10^7$~yr would sweep the galaxy clear of gas. (From King et al.,
      2011)}
\end{figure}

The solutions (\ref{ve}) or (\ref{dotr}) describe the motion of the
contact discontinuity where the shocked wind encounters swept--up
interstellar gas (see Figures 8 and 9). This interface is
strongly Rayleigh--Taylor (RT) unstable, because the shocked wind gas
is highly expanded and has much lower mass density than the
swept--up interstellar gas outside it, so that we have a light fluid underneath a heavy one. 
The RT instability leads to strong overturning motions even on small scales, and so is difficult
to handle numerically. Deductions concerning the mean velocity and energy of the outflow, and its average spatial scale $R(t)$, are likely to be
believable, and agree closely with observations (see below) but we should be very cautious
about results depending strongly on the detailed nature of the interface between
the shocked wind and the swept--up interstellar medium. The RT instability is probably
the reason that the high--speed ($\sim 1000~{\rm km\,s^{-1}}$) outflows with prodigious mass rates we predict here are generally seen with much of the outflowing gas in molecular form. Apparently the interstellar gas entering the forward shock is efficiently cooled by
two--body radiation processes. A preliminary analysis (Zubovas \& King 2014) suggests that
the interstellar gas overtaken by the forward shock is likely to have a multiphase structure. Most
of it cools all the way from the shock temperature $\sim 10^7$~K back to low temperatures, ending in largely molecular form, even though it is entrained in an outflow with the
$\sim 1000~{\rm km\,s^{-1}}$ velocity of the forward shock. But cooling is affected by the topology of the gas flow and the total area of interfaces between
different gas phases. A full numerical calculation of this is currently impossible, so for the time being we can only make comparison with simple estimates, as here.

The mass outflow rate is fixed by how fast the outer shock overtakes the ISM and entrains new interstellar gas ahead of the contact discontinuity. The ISM ahead of the shock is at rest, so this runs into it at a speed giving a velocity jump by a factor $(\gamma +1)/(\gamma -1)$ in the shock frame (where $\gamma$ is the specific heat ratio: see e.g. Dyson \& Williams 1997 for a derivation). This fixes its velocity as
\begin{equation}
v_{\rm out} = {\gamma + 1\over 2}\dot R \simeq
1230\sigma_{200}^{2/3}\left({lf_c\over f_g}\right)^{1/3}~{\rm km\ s}^{-1}
\label{vout}
\end{equation}
(where we have used $\gamma= 5/3$ in the last form, 
and $f_c \simeq 0.16$ is the cosmological value of
$f_g$). This implies a shock temperature of order $10^7$~K for
the forward (ISM) shock (as opposed to $\sim
10^{10 - 11}~K$ for the wind shock). Since the outer shock and the
contact discontinuity were very close together as energy--driven flow
took over from momentum--driven flow (see Figure 8) this means that the outer shock is
at
\begin{equation}
R_{\rm out}(t) = {\gamma + 1\over 2}R(t) = {\gamma + 1\over 2}v_e t.
\end{equation}
This gives the mass outflow rate as
\begin{equation} 
\dot{M}_{\rm out} = {{\rm d}M(R_{\rm out})\over {\rm d}t} = {(\gamma +
  1)f_{\rm g} \sigma^2\over G}\dot R.
\end{equation} 
For comparison the mass rate of the black hole wind, assuming $M = M_{\sigma}$, is
\begin{equation}
\dot{M}_{\rm w} \equiv \dot{m}\dot{M}_{\rm Edd} = \frac{4 f_{\rm c} \dot{m}
  \sigma^4}{\eta c G}.
\end{equation}
This is much smaller than the outflow rate $\mo$ it drives, so 
we define a mass--loading factor as the ratio 
of the mass flow rate in the shocked ISM to that in the wind:
\begin{equation}
f_{\rm L} \equiv {\mo\over\mw} = {\eta(\gamma + 1)\over 4\dot m}{f_g\over
  f_c}{\dot Rc\over \sigma^2}.
\label{load}
\end{equation}
Then we have
\begin{equation}
\mo = f_{\rm L}\mw = {\eta(\gamma + 1)\over 4}{f_g\over f_c}{\dot Rc\over
  \sigma^2}\me.
\end{equation}
If the AGN radiates at a luminosity $\sim \led$, we
have $\dot R = v_e$, and  (\ref{ve}) gives
\begin{equation}
f_{\rm L} = \left({2\eta c\over 3\sigma}\right)^{4/3}\left({f_g\over
  f_c}\right)^{2/3}{l^{1/3}\over \dot m} \simeq
460\sigma_{200}^{-2/3}{l^{1/3}\over \dot m},
\label{load2}
\end{equation}
and
\begin{equation}
\mo \simeq 4060\sigma_{200}^{10/3}l^{1/3}
~\msun\,{\rm yr}^{-1}
\label{out}
\end{equation}
for typical parameters, $f_{\rm g} = f_{\rm c}$ and $\gamma =
5/3$.  The total gas mass in the bulge is roughly $M_g \sim 10^3f_gM_\sigma$ (from equation \ref{mmb}).
Clearly if the outflow persists for a time $t_{\rm clear} \sim M_g/\mo \sim 1\times 10^7 \sigma_{200}^{2/3}l^{-1/3}$~yr
it will sweep away a large fraction of the galaxy's gas. The precise outflow duration needed for this depends on both the type and the environment of the galaxy, in practice leading to three parallel but slightly offset $M - \sigma$ relations (see Section 5.5 below).

Equations (\ref{vout}, \ref{out}) give 
\begin{equation} 
\frac{1}{2}\dot{M}_{\rm w} v^2 \simeq
\frac{1}{2}\dot{M}_{\rm out} v_{\rm out}^2.  
\label{energy2}
\end{equation} 
So most of the wind kinetic energy ultimately goes into the
mechanical energy of the outflow, as we would expect for energy driving. The
continuity relations across the contact discontinuity show that if the
quasar is still active, the shocked wind retains $1/3$ of the total
incident wind kinetic energy $\dot M_{\rm w}v^2/2$, giving
$2/3$ to the swept--up gas represented by $\dot M_{\rm out}$.

Equation (\ref{energy2}) means that the swept--up gas must have a scalar momentum rate
greater than the Eddington value $\led/c$, since we can rewrite it as
\begin{equation} 
\frac{\dot{P}_{\rm w}^2}{2 \dot{M}_{\rm w}} \simeq \frac{\dot{P}_{\rm
    out}^2}{2 \dot{M}_{\rm out}}, 
\end{equation}
where the $\dot P$'s are the momentum fluxes. With $\dot{P}_{\rm w} = L_{\rm
  Edd}/c$, we have
\begin{equation} \label{eq:dotp}
\dot{P}_{\rm out} = \dot{P}_{\rm w} \left(\frac{\dot{M}_{\rm
    out}}{\dot{M}_{\rm w}}\right)^{1/2} = \frac{L_{\rm Edd}}{c} f_{\rm
  L}^{1/2} \simeq 20 {\led\over c}\sigma_{200}^{-1/3}l^{1/6}.
\end{equation} 
Observations of molecular outflows
consistently show $\dot{M}_{\rm out} v_{\rm out} > L_{\rm Edd}/c$, and in particular Cicone et al. (2014) find that momentum rates $20L/c$ are common. This is an inevitable consequence of mass--loading ($f_L > 1$). These high momentum rates are important, 
as they are probably the way that the galaxy resists the accretion that
cosmological simulations suggest still continues at large scales (Costa et al. 2014). 

Recent infrared observations show abundant evidence for molecular outflows with speeds and mass rates similar to (\ref{vout}) and ({\ref{out}).
Feruglio et al. (2010), Rupke \& Veilleux (2011) and Sturm et al. (2011) find large--scale (kpc)
flows with $v_{\rm out} \sim 1000$~km\,s$^{-1}$) and $\mo \sim
1000~\msun\, {\rm yr}^{-1}$) in the nearby quasar
Mrk~231. Other galaxies show similar phenomena (cf
Lonsdale et al. 2003, Tacconi et al. 2002, Veilleux et al. 2009
Riffel \& Storchi--Bergmann (2011a, b) and Sturm et al. 2011: see Tables 1 and 2 of Zubovas \& King 2012a for a detailed comparison with the theoretical predictions). In each case it
appears that AGN feedback is the driving agency. There is general agreement 
for Mrk231 for example that the mass outflow rate $\dot
M_{\rm out}$ and the kinetic energy rate $\dot E_{\rm out} = \dot
M_{\rm out}v_{\rm out}^2/2$ are too large to be driven
by star formation, but comparable with values predicted for AGN feedback. 

It appears that energy--driven outflows from SMBH which have just reached
their $M - \sigma$ masses should be able to sweep galaxy spheroids clear of
gas. A robust observational test of
this is the expected mechanical luminosity (cf equation \ref{lum})
\begin{equation}
L_{\rm mech} \sim {\eta\over 2}\led \simeq 0.05L,
\label{mechout}
\end{equation}
where $L = l\led$ is the observed AGN luminosity. This is investigated by Cicone et al. (2014).
As their Figure 12 shows, observation does largely confirm the relation {\ref{mechout}). 
If the AGN are close to their Eddington luminosities (so that  $L \propto M \propto \sigma^4$ and $l\simeq 1$), the clearout rate $\propto \sigma^{10/3}$ (equation \ref{out}) should scale linearly with the driving luminosity $L$.  Figure 9 of Cicone et al. shows evidence for  this correlation, with normalization close to that predicted.

\subsection{Effects of a Galaxy Disc: Stimulated Star Formation and Outflow Morphology}

We have so far discussed galaxy spheroids in isolation. This is in line with the observational evidence (see Kormendy \& Ho 2013 for a review) that the SMBH scaling relations apply only to this component of a galaxy, and are essentially unaffected by the presence of a galaxy disc. In particular we suggest that the critical $M - \sigma$ black hole mass is set by small--scale momentum--driven outflows interacting with only a very small central part of the bulge. 
But the energy--driven outflows we considered in the last subsection are global: they expand to far greater scales, and unless the galaxy is an elliptical must inevitably encounter its disc as they expand. In a gas--rich galaxy the gas in the innermost disc at radius $R_0$ must be close to self--gravitating. Assuming that the potential is roughly isothermal, it is straightforward to show that this implies a gas density $\sim \rho_d \sim 2\sigma^2/R_0$, i.e. greater than the bulge gas density by the factor $\sim 1/f_g \sim 10$. We see from equation (\ref{motion4}) that higher gas densities mean lower spherical outflow velocities, as they meet greater resistance. So when an initially spherical outflow encounters a high--density gas disc it flows around it, over its plane upper and lower faces. But the pressure in the outflow is at least initially far higher than in the disc: we can read off the pressure at the contact discontinuity from equation (\ref{motion2}) as 
\begin{equation}
P_{\rm CD} = {f_g\sigma^2(2\sigma^2 + v_e^2)\over \pi  GR^2} \simeq {f_g\sigma^2v_e^2\over \pi GR^2}
\label{pcd}
\end{equation}
and estimate the pressure at the forward shock into the ISM as
\begin{equation}
P_{\rm fs} = {4\over 3}\rho(R)v_e^2 = {2f_g\sigma^2v_e^2\over 3\pi GR^2} \simeq {2\over 3}P_{\rm CD}.
\label{pfs}
\end{equation}
By contrast the mid--plane pressure in a disc close to self--gravitating is
\begin{equation}
P_{\rm disc} \sim \rho c_s^2 \sim \rho \sigma^2 \sim 2{\sigma^4\over GR_d^2}
\label{pdisc}
\end{equation}
where we have assumed the sound speed $c_s \sim \sigma$ and the self--gravity condition $G\rho \sim \Omega^2$ with $\Omega = \sqrt{2}\sigma/R_d$ the Kepler frequency at disc radius $R_d$. Thus when the outflow shock arrives at $R = R_d$ its pressure is a factor $\sim (v_e/\sigma)^2 \sim 25$ larger than the disc's, and this remains true until the outflow shock has travelled out to radii $R > R_dv_e/\sigma \sim 5R_d$. Any such compression must trigger a burst of star formation in the disc (cf Thompson et al. 2005, Appendix B), and here it rises to values 
\begin{equation}
\dot \Sigma_{\ast} \sim 2000 \epsilon_{-3}\sigma_{200}^{10/3}l^{2/3}R_{\rm kpc}^{-2}~\msun\, {\rm kpc}^{-2}
\label{sfr}
\end{equation}
(Zubovas et al. 2013), where $\epsilon_{\ast} = 10^{-3}\epsilon_{-3}$ is the efficiency of massive stars in converting mass into radiation, and we have substituted for $v_e$ using (\ref{ve}). Zubovas et al. (2013) show that this leads to a starburst
of total luminosity
\begin{equation}
L_\ast \simeq 5\times 10^{47}L_{46}^{5/6}l^{-1/6}~{\rm erg\,s^{-1}},
\label{starburst}
\end{equation}
where $L_{46}$ is the AGN luminosity in units of $10^{46}~{\rm erg\,s}^{-1}$. Such systems would appear as ULIRGs. 

This suggests that in a galaxy with both a bulge and a disc, the clearout phase leaves the galaxy bulge without gas, but may be accompanied by a starburst in the disc. Recent observations of dusty QSOs appear to show this, with the black hole mass already on the $M - M_b$ relation (\ref{mmb}), and so fully grown (Bongiorno et al. 2014). In an elliptical on the other hand, clearout must leave the galaxy `red and dead'.

Since a galaxy disc is a major obstacle to an outflow, it follows that it may be able to divert a quasi--spherical outflow into a bipolar shape. This is particularly true in cases where the SMBH mass grows only a little, in a minor accretion event. Zubovas et al. (2011) suggest that the gamma--ray emitting bubbles disposed symmetrically about the plane of the Milky Way (Su et al. 2010) may be the remnants of a relatively recent and rather weak event like this.

\subsection{The Three $M - \sigma$ relations}

So far in this Section we have seen that the arrival of the black hole mass
at the $M - \sigma$ relation means that its feedback makes a radical change
from momentum--driving to energy-driving. The 
energy--driven phase which clears the gas from a galaxy bulge is short and 
violent. But it is clear that the black hole must inject a non--negligible amount
of energy to eject the gas, and this requires accretion energy, i.e. some black hole
mass growth. Evidently if the mass increment $\Delta M$ needed for this is
$\gg M_{\sigma}$ we will have failed to explain the $M - \sigma$ relation. 

\begin{figure}
  \centering
    \includegraphics[width=\textwidth]  {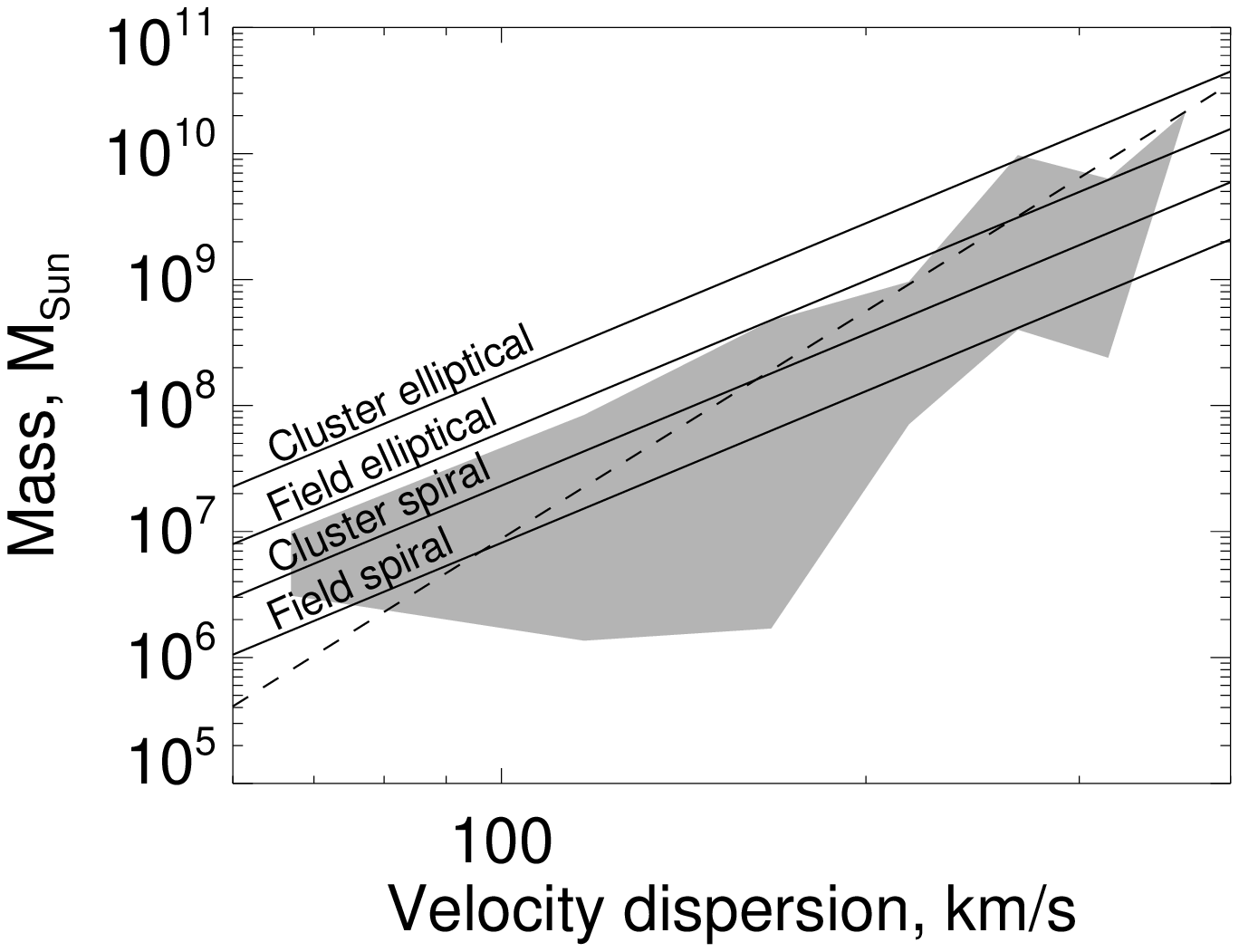}
    \caption{The four (in reality three, as cluster spirals are rare)
      $M-\sigma$ relations (solid lines) and their combined effect on
      observational fits (dashed line). All solid lines have slopes $M
      \propto \sigma^4$ and the dashed line has $M \propto
      \sigma^6$. The grey area is the approximate locus of data points
      in Figure 3 of McConnell et al. (2011). (From Zubovas \& King,
      2012b)}
\end{figure}

The mass increment $\Delta M$ is influenced by two factors.
First, it must require significantly less SMBH growth to remove the gas from
a spiral galaxy with a relatively small bulge, than for example an elliptical, where 
the much larger bulge mass means that energy--driving by
the central SMBH wind must continue for longer in order to expel the
remaining gas.  Zubovas \& King (2012b) find that energy--driving, and
therefore SMBH mass growth above $M_{\sigma}$, must continue only for about 4 Myr (about 0.1 Salpeter times) in a typical spiral, but for
for about 2 Salpeter times in an elliptical. So the final SMBH mass in a spiral is close to $M_{\sigma}$ but in an elliptical it can reach
\begin{equation}
M_{\rm final} \sim e^2 M_{\sigma} \sim 7.5M_{\sigma}
\label{finmass}
\end{equation}
\begin{figure}
  \centering
    \includegraphics[width=\textwidth]  {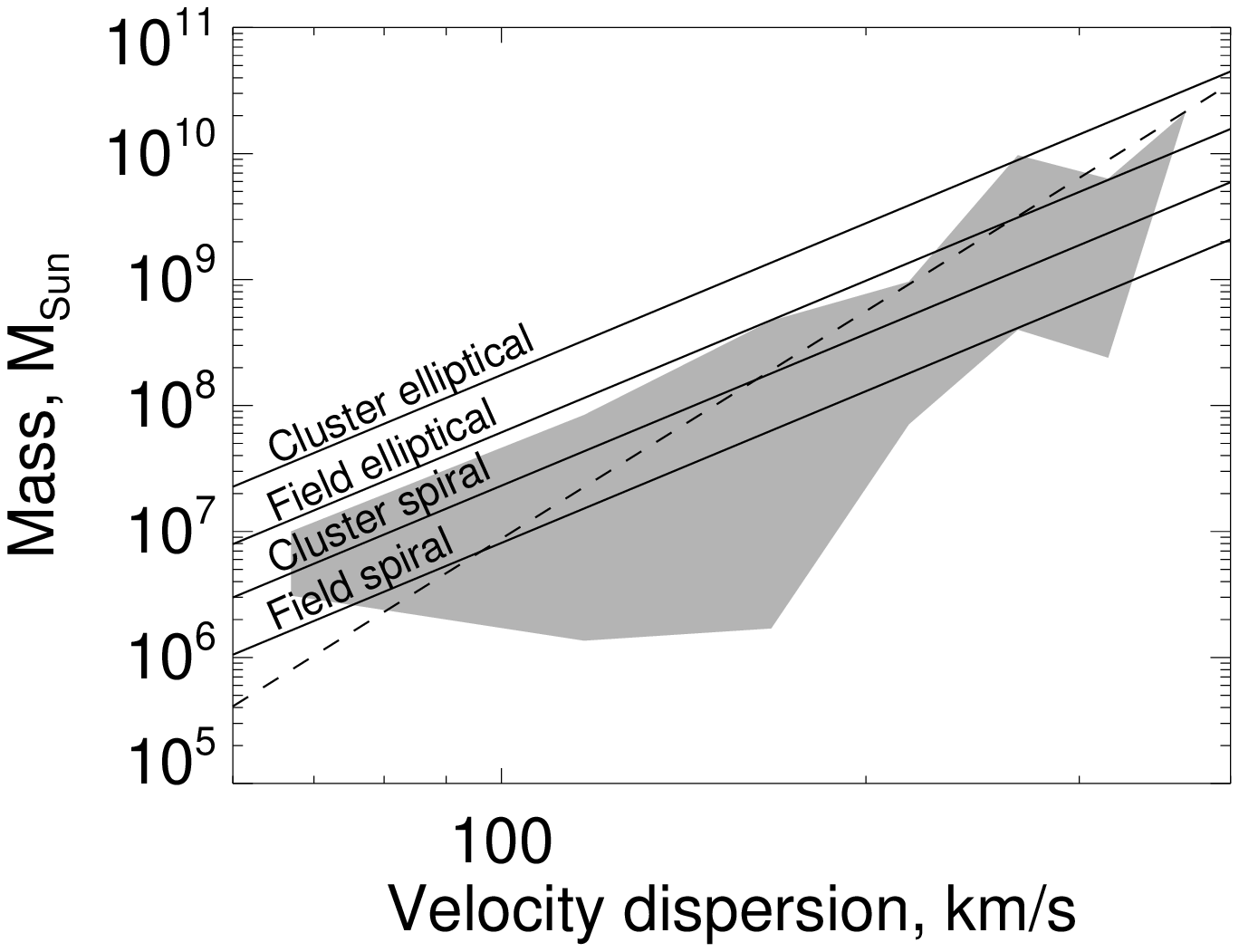}
    \caption{The four (in reality three, as cluster spirals are rare)
      $M-\sigma$ relations (solid lines) and their combined effect on
      observational fits (dashed line). All solid lines have slopes $M
      \propto \sigma^4$ and the dashed line has $M \propto
      \sigma^6$. The grey area is the approximate locus of data points
      in Figure 3 of McConnell et al. (2011). (From Zubovas \& King,
      2012b)}
\end{figure}
The second factor affecting $M, M_b$ is the galaxy environment. 
Cluster ellipticals can gain gas as they orbit through the intracluster gas. Some Brightest Cluster
Galaxies (BCGs), which are near the centre of the cluster potential,
are known to contain unusually massive SMBH (McConnell
et al. 2011). Taking account of the extra black hole mass growth required to remove 
the bulge, and the mass a galaxy may gain from its surroundings, implies three parallel but
slightly offset $M - \sigma$ relations for spirals, field and cluster
ellipticals (see Fig 10). In principle there is also
a relation for cluster spirals, but these are rare.
We see from the Figure that the spread in offsets means that an
observed sample drawn from galaxies of all three types would
tend to produce a slope slightly bigger than the individual ones for each type, perhaps
accounting for the slight discrepancy between the observed overall slope $\alpha = 4.4\pm 0.3$
and the theoretical value of 4. All
three types of galaxies obey a similar $M - M_b$ relation (\ref{mmb}) within
the errors, as growth of the SMBH above $M_{\sigma}$ goes together with
higher $M_b$. 


\section{THE SMBH -- BULGE MASS RELATION}

\subsection{Feedback and the $M - M_b$ relation}

In the Introduction we noted the observed  proportionality (\ref{mmb})  between $M$ and $M_b$ as well as the $M - \sigma$ relation. So far we have concentrated almost entirely on the second of these relations, and suggested that it arises because the black hole feedback itself directly limits the mass reservoir available for black--hole growth. Quite independently of details, almost every discussion of this relation adopts this view (see Section 7 below). 

But the character of the $M - M_b$ relation must be very different. Since we are assuming that feedback ensures that the black hole mass $M$ is set by 
$\sigma$ we cannot argue that $M$ is independently set by $M_b$. But reversing the argument to suggest that the black hole mass $M$ sets $M_b$ is also implausible, since $M_b$ is in the form of stars. 


So there can be no directly causal connection between the black hole mass $M$ and the stellar bulge mass $M_b$. (Indeed one view -- see Section 6.3 below -- asserts that the connection is purely statistical.) Instead, their relation must arise because whatever determines $M_b$ makes it proportional to $\sigma^4$. Empirically, we already know that this is approximately true, at least for elliptical galaxies, the largest spheroids of all, because these are observed to obey the Faber--Jackson (1976) relation 
\begin{equation}
L_{\ast} \sim 2\times 10^{10}\lsun\sigma_{200}^4,
\label{fj}
\end{equation}
Here $L_*$ is the total stellar luminosity and mass of an elliptical, so for mass--to--light ratios $\sim 5$ we immediately get the stellar mass as
\begin{equation}
M_* \sim 1\times 10^{11}\msun\sigma_{200}^4 \sim 10^3 M
\label{fjm}
\end{equation}
There is now general agreement that this relation, like the $M-\sigma$ relation,  may result from feedback inhibiting and ultimately suppressing the process that produces it. The difference is that here the feedback is from stars, and what ultimately has to be suppressed is star formation. 
Several papers make this point, starting with Murray et al. (2005). 
%
Power et al. (2011) show that this approach gives a bulge stellar mass
\begin{equation}
M_b \sim {0.14f_gt_H\sigma^4\over \epsilon_{\ast}cG},
\label{bulge}
\end{equation}
where $\epsilon{\ast} \simeq 2\times 10^{-3}$ measures the total luminous energy yield from a main--sequence star in terms of its rest--mass energy $M_{\ast}c^2$, and $t_H$ is the Hubble time.
Comparing with (\ref{sig}) we get
\begin{equation}
M \simeq M_{\sigma} \sim {1.8\kappa\epsilon_{\ast}c\over \pi Gt_H}M_b
\sim 10^{-3}M_b,
\label{mmb2}
\end{equation}
which is similar to observational estimates (cf equation \ref{mmb}). Both the $M - \sigma$ and the $M - M_b$ relations hold for elliptical galaxies, so equation (\ref{bulge}) automatically reproduces the Faber--Jackson (1976) relation for typical mass--to--light ratios. In this view, the similarity of the SMBH and stellar (Faber--Jackson) $M, M_b \propto \sigma^4$ relations  (\ref{msig}, \ref{fjm})  follows directly because both result from momentum feedback, and the ratio $M/M_B \sim 10^{-3}$ reflects the relative efficiencies of the black hole and stellar versions.

\subsection{The $M - \sigma$  relation for Nucleated Galaxies}

A similar argument (McLaughlin et al. 2006)
shows that for nucleated galaxies (i.e those whose central regions
are dominated by nuclear star clusters, with no detectable sign of the
presence of a supermassive black hole) there should be an offset $M - \sigma$
relation between the mass of the cluster and the velocity dispersion, i.e.
\begin{equation}
M_c \simeq 20M_{\sigma} \simeq 6\times 10^9\sigma_{200}^4\msun.
\label{clustermsig}
\end{equation}
Typically these galaxies are small, with $\sigma < 120~{\rm
  km\,s^{-1}}$. The factor  $\sim 20$ offset in cluster mass for a
given $\sigma$ arises because momentum--driving by 
an ensemble of cluster stars  is about 20 times less efficient per
unit mass than from a black hole accreting at the Eddington rate.

\subsection{Mergers and the $M - M_b$ relation}

Jahnke \& Macci\`o (2011) offer a radically different interpretation of the $M - M_b$ relation. Building on earlier work of Peng (2007) they assume that black hole and bulge masses are built up by repeated mergers of smaller galaxies with uncorrelated $M$ and $M_b$. They follow this evolution using dark matter halo merger trees, and 
as a result of the central limit theorem
find that $M$ is roughly  proportional to $M_b$, with the scatter decreasing for larger masses, where there have been more mergers. They conclude that the SMBH - bulge scaling relation may have an explanation that is largely or even entirely non--causal.


But it is hard to accept that there is no more physics in the SMBH scaling relations than this. First, the actual ratio $M/M_b$ is left undetermined by this procedure. Second, to get from the $M - M_b$ relation to $M - \sigma$ requires one to assume something like the $M_b\propto \sigma^4$ relation (\ref{fjm}) implied by Faber--Jackson, so physics presumably must enter here too (cf the subsection above). Finally, it would seem a remarkable coincidence that the outcome of this indirect process by chance produces an $M - \sigma$ relation exactly equivalent to requiring that the SMBH Eddington thrust should just balance the weight of the bulge gas.

\section{MOMENTUM, ENERGY OR RADIATION?}

The study of AGN outflows and their effects on the host galaxy has two main aims. A viable picture must explain both the scaling relations, and simultaneously the fact that galaxy spheroids appear ultimately to be largely swept clear of gas by high--speed molecular outflows which have significantly greater scalar momenta $\dot M v$ than the radiative value $L/c$ of the central AGN (the clearout problem). The discussion given above offers plausible physical grounds that the shock interaction characterising the black hole wind feedback changes from momentum--driven, acting on small spatial scales near the black hole, to energy--driven, instead acting globally on the whole galaxy bulge and producing a high--energy clearout of its gas. The $M - \sigma$ relation marks the point where this transition occurs in a given galaxy. 
We will argue below (Section 8) that observations support this picture of local--global transition in several ways, but before accepting this conclusion we should consider other possibilities.

First, the switch from momentum to energy--driving depends on the details of gas cooling. It is sometimes argued (e.g. Silk \& Nusser 2010) that strong cooling of the ambient interstellar medium enforces momentum--driving by the central SMBH throughout. In fact cooling the ambient gas is not relevant to the question of energy or momentum driving: as we have seen, it is the cooling of the shocked wind gas which decides this. But as we emphasised in Section 4, at least some of the physics of the suggested momentum--energy switch is still beyond a full numerical treatment with realistic parameters. It is sensible then to check our treatment above by considering the momentum--driven and energy--driven cases in isolation,
and then the effect of direct radiation pressure.


\subsection{Wind Momentum Driving}

We first simply assume that a black hole wind acts on its surroundings by pure momentum--driving alone, at all radii. For generality we let the pre--shock wind have speed $v_w$ and take its mechanical luminosity $\dot M_wv_w^2/2$ as a fixed fraction $a$ of $\led$, i.e. we do not explicitly assume that the wind has the Eddington momentum, as seems to hold for UFOs. Then the momentum feedback first becomes important at a critical black hole mass $M_{\rm crit}$ roughly given by equating the wind thrust $\dot M_w v_w = 2a\led/v$ to the weight
\begin{equation}
W = {GM(R)M_g(R)\over R^2} = {4f_g\sigma^4\over G}
\label{weight}
\end{equation}
of the overlying gas in an isothermal potential (cf equation \ref{motion1}). With $\led = 4GM_{\rm crit}c/\kappa$ we get
\begin{equation}
M_{\rm crit} = {v_w\over 2ac}M_{\sigma}.
\label{fab2}
\end{equation}
By definition $a = \dot M_wv_w^2/2\led$ and $\led =
\eta\me c^2$, so  
\begin{equation}
M_{\rm crit} = {v_w\over 2ac}M_{\sigma} = {\eta c\over v_w}{\me\over
  \dot M_w}M_{\sigma},
\label{fab}
\end{equation}
(cf Fabian 1999).
We see that for general wind parameters the critical mass differs from $M_\sigma$. We find  $M_{\rm crit} = M_{\sigma}$ only if $v_w = \eta c \me/\dot M_w$, which is equation (\ref{v}). This immediately implies that the wind momentum is Eddington, i.e. $\dot M_wv_w = \eta\me c = \led/c$. In other words, assuming pure momentum driving gives the critical mass as $M_{\sigma}$ if and only if the driving wind has the Eddington momentum, i.e. has the properties observed for UFOs. 

But 
pure momentum driving is unable to drive off the bulge gas without a significant increase of the black hole mass above $M_{\rm crit}$. Several authors have reached this conclusion (cf Silk \& Nusser 2010,  
McQuillin \& McLaughlin 2012). 
Moreover,
if galaxy bulges accrete at the rates suggested by cosmological simulations it seems unlikely that any hypothetical momentum--driven outflows would have enough thrust to prevent infall and so could not turn off 
star formation definitively (cf Costa et al. 2014). We conclude that pure momentum--driving, even given the lack of a likely shock cooling process, probably does not give a realistic picture of the interaction 
between SMBH and their hosts.

\subsection{Wind Energy Driving}

The direct opposite case from that considered in the last subsection is pure energy--driving by winds, where radiative cooling is assumed negligible throughout. This was often the implicit assumption in early 
treatments (e.g. Silk \& Rees 1998, Haehnelt et al. 1998). The equation of motion for this case is (\ref{motion4}). This shows that gas is always driven out at constant speed for any SMBH mass, however small: 
setting $R = v_et$ and using the definition
of $\led$ (cf equation \ref{edd}) gives the speed $v_e$ as
\begin{equation}
v_e^3 = {\pi G^2c\eta M\over 3f_g\kappa\sigma^2}. 
\end{equation}
This expresses the fact that the adiabatically expanding shocked gas pushes the interstellar gas away as a hot atmosphere for any SMBH mass. One can easily find the corresponding mass outflow rate by setting 
$\dot M_{\rm out}v_e^2/2 \sim L_w$, since we know that the outflow mechanical luminosity is a significant fraction of the driving wind mechanical luminosity $L_w = \eta l\led/2$.

To define a critical SMBH mass for energy--driven outflow one has to impose a further condition. This is usually taken as $v_e \sim \sigma$, defining some kind of escape velocity. But it is not obvious that 
this is appropriate: the outflow is driven by pressure, so the ballistic escape velocity is not relevant. Even if the AGN switches off, the residual gas pressure still drives outflow for a long time (cf Fig. 9). 
If we nevertheless impose this condition we find a critical mass
\begin{equation}
M_{\rm energy} = {3f_g\kappa\over \pi G^2 \eta c}\sigma^5 = {3\sigma\over \eta
  c}M_{\sigma} = 0.02M_{\sigma} = 6\times 10^6\msun\sigma_{200}^5
\label{menergy}
\end{equation}
which is a factor $3\sigma/\eta c\sim 1/50$ too small in comparison with observations. 


Silk \& Rees (1998) considered the growth of a protogalaxy (i.e. gas with $f_g \sim 1$) around a supermassive seed black hole which formed earlier, but their argument applies to the coevolution of the SMBH and 
host also, provided we take $f_g \sim 0.1$. They assume the wind sweeps mass into a shell with speed $v_s$, and implicitly neglect pressure work, and the fact that energy is shared between the shocked wind and 
the swept--up outflow. This would imply a relation
\begin{equation}
L_w = 2\pi r^2f_g\rho(r) v_s^3,
\end{equation}
as each new shell of mass $4\pi r^2\rho(r) v_s$ now simply acquires kinetic energy $v_s^2/2$ as it is swept up. Using the isothermal relation
(\ref{rho}) and requiring $v_s \sim \sigma$ gives
\begin{equation}
M_{\rm SR} \simeq {f_g\kappa\over 4\pi G^2 f_wc}\sigma^5 \simeq 5\times 10^4\left({f_g\over 0.16}\right)\msun\sigma_{200}^5
\label{SR}
\end{equation}
where $f_w = L_w/\led$. The neglect of pressure work overestimates the wind--driving efficiency, so this mass is even smaller than (\ref{menergy}). It is clear that wind energy--driving of this type does not 
correctly reproduce the $M - \sigma$ relation, giving a critical mass too low by factors 50 -- 100.

A more promising approach has recently been outlined by Nayakshin (2014), Zubovas \& Nayakshin (2014) and Bourne et al. (2014), who consider the effects of strong inhomogeneity of the bulge gas. They assume 
first that inverse Compton shock cooling may not be effective because of two--temperature effects (but see Section 4.2 above). Second, they suggest that sufficiently dense clouds of interstellar gas would 
feel a net outward force $\sim \rho v^2$ per unit area when overtaken by a free--streaming UFO wind of preshock density $\rho$ and speed $v$, thus mimicking a momentum--driven case. The density of these 
clouds is a factor $1/f_g \sim 6$ below the star--formation threshold. If most of the ISM gas mass is in the form of such clouds, SMBH feeding must stop when the outward force overcomes gravity, which 
gives a relation like (\ref{sig}) up to some numerical factor. This idea throws up several gas--dynamical problems. First, a cogent treatment must explain the origin and survival of clouds at densities 
close to but just below the star formation threshold, which must contain most of the interstellar gas. The clouds must be completely immersed in the wind, so the net outward force on them is a surface 
drag, which is dimensionally also $\sim \rho v^2$ per unit area. Estimating this surface drag requires knowledge of how the cloud--wind interfaces evolve on very small scales. 
Since these are formidable theoretical tasks, we should ask for observational tests. The main difference from the quasispherical momentum--driven case is that instead of being radiated away,  most of the 
energy of the UFO wind now continuously drives the tenuous intercloud part of the ISM out of the galaxy at high speed. If this tenuous gas is a fraction $f_t$ of the total gas content, equations 
(\ref{vout}, \ref{out}) show that for SMBH masses not too far from $M_{\sigma}$ this outflow should have speed
$v_{\rm out} \sim 1230 f_t^{-1/3}~{\rm km\, s^{-1}}$
and mass--loss rate $M_{\rm out} \sim 4000f_t~\msun\,{\rm yr}^{-1}$, 
and so be potentially observable for many AGN spheroids. From the work of Section 5.4 
one might also expect a continuously elevated star formation rate in the central parts of their galaxy discs also, which is not in general observed.

\subsection{Cosmological Simulations}

Cosmological simulations often produce an empirical $M - \sigma$ relation as
part of much larger structure formation calculations. Limits on
numerical resolution inevitably require a much more broad--brush
approach then adopted here. The effect of the SMBH on its surroundings is usually
modelled by distributing energy over the gas of the numerical `galaxy' at a certain rate. 
This injected mechanical luminosity is then iterated until the right relation appears. This
empirical approach (e.g. di Matteo et al. 2005) seems always to
require a mechanical luminosity $0.05\led$ to produce the
observed $M - \sigma$ relation. This is precisely what we expect
(cf equation \ref{lum}) for a black hole wind with the Eddington momentum $\mo v =
\led/c$. 

But the success of this procedure is puzzling. If the ambient gas absorbed the full numerically injected mechanical luminosity $0.05\led$ the resulting outflows would
give the energy--driven (\ref{energy}) or Silk--Rees mass (\ref{SR}) above, which are too
small compared with observations. The fact that cosmological simulations instead actually iterate roughly to the observed $M - \sigma$ value (\ref{msig2}) suggests that they somehow
arrange that the injected energy only couples to the gas at
the very inefficient rate which occurs in momentum driving, or possibly that the 
numerical gas distribution is highly inhomogenous. The real
physics determining this in both cases operates at lengthscales far below the
resolution of any conceivable cosmological simulation, so the
inefficiency must be implicit in some of the `sub--grid' physics which
all such simulations have to assume (cf Costa et al. 2014, Appendix B).

\subsection{Radiation Driving}

\subsubsection{Electron scattering opacity}

We remarked in the Introduction that in principle direct radiation pressure is the strongest  perturbation that a black hole makes on its surroundings, but its effects are more limited than this suggests. As we already suggested, the reason is that in many situations radiation decouples from matter before it has transferred significant energy or momentum. This is particularly likely for  radiation emitted by an AGN in the center of a galaxy bulge. The gas density (cf equation \ref{rho}) is sharply peaked towards the center, and
so is its tendency to absorb or scatter the radiation from the accreting black hole. The electron scattering optical depth from gas outside a radius $R$ for example is
\begin{equation}
\tau(R) = \int_R^{\infty} \kappa \rho(r) {\rm d}r = {\kappa
  f_g\sigma^2\over 2\pi GR}\ ,
\label{tauR1}
\end{equation}
which is mostly concentrated near the inner radius $R$. This means that gas initially close to
the AGN is probably swept into a thin shell by its radiation, and so at radius $R$ has optical depth
\begin{equation} 
\tau_{\rm sh}(R) \simeq {\kappa
  f_g\sigma^2\over \pi GR}\ ,
\label{taush}  
\end{equation}
very similar to the undisturbed gas outside $R$ (cf equation \ref{tauR1}).
Gas distributed in this way has large optical depth near the black hole when its inner edge $R$ is
small (i.e. less than the value $R_{\rm tr}$ specified in equation \ref{rtr} below). Then the accumulating accretion luminosity $L$ of
the AGN is initially largely trapped and isotropized by electron scattering, producing a blackbody 
radiation field whose pressure grows as the central AGN radiates.
This growing pressure pushes against the weight $W =4f_g\sigma^4/G$ (equation \ref{weight})
of the swept--up gas shell at radius $R$. This is exactly like the material energy--driving we discussed in Section 5.3, except here the photon `gas' has $\gamma = 4/3$ rather than $\gamma = 5/3$ there. Clearly the effectiveness of this radiation driving
is eventually limited because the shell's optical depth falls
off like $1/R$ as it expands. The force exerted by the radiation drops
as it begins to leak out of the cavity, until for some value $\tau_{\rm tot}(R)
\sim 1$ it cannot drive the shell further.

This shows that the sweeping up of gas by radiation pressure must stop
at a `transparency radius'
\begin{equation}
R_{\rm tr} \sim {\kappa f_g\sigma^2\over \pi G} \simeq
50\left({f_g\over 0.16}\right)\sigma_{200}^2~{\rm pc},
\label{rtr}
\end{equation}
where (up to a logarithmic factor) the optical depth $\tau_{\rm tot}$
is of order 1, so that the radiation just escapes, acting as a safety
valve for the otherwise growing radiation pressure. This process is discussed in in detail by King 
\& Pounds (2014), who suggest that the stalled gas at $R_{\rm tr}$ may be the origin of
the `warm absorber' phenomenon (cf Tombesi et al. 2013). The radius $R_{\rm tr}$ is so small that very little accretion energy is needed to blow interstellar gas to establish this structure, and to adjust it as the galaxy grows and changes $\sigma$.

\subsubsection{Dust}
At larger radii much of the cold diffuse matter in the galaxy bulge may be in the form of dust. The absorption coefficient of dust depends strongly on wavelength and is far higher than electron scattering in the ultraviolet, but decreases sharply in the infrared (e.g. Draine \& Lee, 1984). The energy of an ultraviolet photon absorbed by a dust grain may be re--emitted almost isotropically as many infrared photons, which then escape freely. The net effect is that dusty gas feels only the initial momentum of the incident UV photon, while most of the incident energy escapes. Then a spherical shell around an AGN would experience a radial force $\simeq L/c$, where $L$ is the ultraviolet luminosity, as long as it remained optically thick to this kind of dust absorption. This is dynamically similar to wind--powered flows in the momentum--driven limit, and this type of radiation--powered flow is often also called `momentum--driven', even though the physical mechanism is very different. 

An important distinction between the wind and radiation--powered cases is that 
ambient gas in the path of a wind cannot avoid feeling its effects, whereas this is not true for 
radiation, as the gas may be optically thin. Galaxies are generally optically thin to photons in 
various wavelength ranges, and a radiation--driven shell may stall at finite radius because its optical depth $\tau$ becomes so small that the radiation force decouples, as we saw in the electron--scattering case. Ishibashi \& 
Fabian (2012, 2013, 2014) appeal to this property to suggest that star formation in massive galaxies 
proceeds from inside to outside as radiation--momentum driven shells of dusty gas are driven out and then stall at the dust transparency radius $R_{\rm dust} \simeq (\kappa_d/\kappa) R_{\rm tr}$. For large dust opacities $\kappa_d \sim 1000\kappa$ this can give $R_{\rm dust} \sim  50$~kpc. In contrast galaxies are probably never `optically thin' to winds, and the density of a black hole wind is always diluted as $1/R^2$, so it inevitably shocks against a swept--up shell of interstellar gas at large $R$. 

The mathematical similarity (cf eq \ref{motion1}) between wind--powered and 
radiation--powered momentum driving allows an empirical estimate of an $M - \sigma$ 
relation for the latter if we assume that observed AGN define the relation, and that their observed luminosities correspond to $L/\led \sim 0.1 - 1$. This gives $M_{\rm crit} = (\led/L)M_\sigma \sim 1- 10 M_\sigma$  (Murray et al. 2005). Optical depth effects might narrow this range closer to the observed one (Debuhr et al. 2012). This suggests that radiation driving might be compatible with the $M - \sigma$ relation, but a momentum--driven outflow like this can never simultaneously reproduce the high--velocity molecular outflows characterising the clearout phase. In particular its momentum is $L/c < \led/c$, considerably smaller than the observed 
$\sim 20\led/c$ of such flows (see Section 5.3). In other words, we have the usual difficulty that momentum--driving can accommodate the $M - \sigma$ relation, but not simultaneously solve the clearout problem.

One way of possibly overcoming this (e.g. Faucher--Gigu\`ere et al. 2012; Faucher--Gigu\`ere \& Quataert, 2012) is to assume (cf Roth et al. 2012) that instead of degrading incident high--energy photons to lower--energy ones that escape freely, the effect of dust absorption is to retain much of the incident radiant energy. Then
 if the dust is distributed spherically and is in a 
 steady state the radiation force on it is $\tau L/c$, where $\tau$ is the radial optical depth of 
 the dust (cf Roth et al. 2012).  This form of radiation driving of optically thick dust can in principle produce outflows whose scalar
momenta are boosted above that of the driving luminosity $L/c$ by a factor $\sim \tau$ 
because photons may be reabsorbed several times. For $\tau \gg 1$ the radiation field is effectively 
trapped and presumably approaches a blackbody form (cf the discussion of the electron 
scattering case above), limiting the boost. 

Evidently for radiation driving of dust to solve simultaneously both the SMBH scaling and clearout problems requires a sharp transition in the properties of the dust opacity at the critical $M - \sigma$ mass. This must change the outflows from effectively momentum--driven (incident photons are absorbed but their energy escapes as softer photons)
 to energy--driven (incident photons trapped) at this point, in a switch analogous to the turnoff of Compton cooling in the wind--driven case. There have so far been no suggestions of how this might happen, but the physics of dust opacity is sufficiently complex that this is perhaps not surprising.


\section{SMALL vs LARGE SCALE FEEDBACK}

We have shown that UFO winds are very common in AGN, despite quite restrictive conditions on their observability. They provide an obvious way for the central supermassive black hole to communicate its 
presence to its host. This in turn suggests ways of understanding both the SMBH -- galaxy scaling relations, and the need to expel gas from the galaxy spheroid to terminate star formation. It is clear that this AGN feedback must operate at times on small scales, and at others on large scales. Our discussion of feedback points to a natural association between momentum--driving and small scales, and between energy--driving and large scales. 
Small--scale phenomena naturally explained by wind momentum--driving include

1. Super--solar elemental abundances in AGN spectra. Wind momentum--driving automatically sweeps up and compresses the same gas many times before the black hole mass reaches $M_\sigma$. Generations of 
massive stars forming out of the same swept--up gas can repeatedly enrich the gas close to the SMBH with nuclear--processed material before the $M_\sigma$ mass is reached, and momentum--driving changes 
to energy--driving. 

2. Dark matter cusp removal. The same repeated sweeping--up of a gas mass comparable to the SMBH mass, followed by fallback, has a strong tendency to weaken dark matter cusps. Because the baryonic mass 
involved is much larger, this is a more powerful version of the mechanism invoked by Pontzen \& Governato (2012)  (see also Garrison--Kimmel et al. 2013), who considered supernovae near the SMBH.

3. Quiescence of AGN hosts. Most AGN hosts do not show dramatically elevated star formation in the central regions of their galaxy discs, or so far much evidence for high--speed  
($\sim 1000~{\rm km\, s^{-1}}$ and massive ($\sim {\rm few}\,100\msun\,{\rm yr}^{-1}$) outflows on large scales. This is compatible with wind driving by momentum but not energy.

Large--scale phenomena suggesting the action of energy--driving include

4. Metals in the circumgalactic medium. These must be made in galaxies and only later expelled to make the CGM. This suggests that expulsion through energy--driving acts only after stellar evolution has 
had time to enrich a significant fraction of the galaxy bulge gas.

5. Mechanical luminosities of galaxy--scale molecular outflows. These are observed to be close to 5\% of their central AGN luminosities $L$, just as expected for energy--driving, with momenta close to 
the predicted $20L/c$.

6. Suppression of cosmological infall. Energy--driven outflows at large radii probably prevent galaxies accreting indefinitely (Costa et al. 2014).

This list seems to favor a combination of momentum and energy driving, with some kind of switch between them. The suppression of inverse Compton shock cooling at the point when the black hole mass 
reaches $M_\sigma$ appears promising, but requires further work on how observable the cooling is, and the possibility of two--fluid effects. Radiation driving on dust could produce similar behavior, 
although the physics controlling the required switch between momentum and energy driving is so far unexplained.

It is worth stressing that even a detailed understanding of the dual role of AGN feedback in establishing both the SMBH -- host scaling relations and the quenching of star formation would solve only 
half of the problem. 
For a full picture of how black holes and galaxies influence each other
we need to know what physical mechanism can produce a supply of gas with so little angular momentum that much of it can accrete on to the central supermassive black hole within a few Salpeter times 
(see Section 3.1, and equation \ref{tvisc1}). We saw in Section 1.2 that the hole's gravity is far too weak to influence the galaxy on the mass scale needed for this. Only feedback can do this, perhaps suggesting that SMBH feedback may ultimately cause SMBH feeding (cf Dehnen \& King 2013). 

\noindent{\bf ACKNOWLEDGMENTS}

We thank Walter Dehnen, Dean McLaughlin, Sergei Nayakshin, Chris Nixon, Chris Power,  Jim Pringle,  James Reeves, Simon Vaughan, Mark Wilkinson and Kastytis Zubovas for help, collaboration and advice on 
the subjects reviewed here. We have benefitted hugely from discussions with many people over the years, including Mitch Begelman, Martin Elvis, Andy Fabian, Claude--Andr\'e Faucher--Gigu\`ere, Reinhard 
Genzel, Martin Haehnelt, Luis Ho, Knud Jahnke, Roberto Maiolino, David Merritt, Ramesh Narayan, Ken Ohsuga, Brad Peterson, Eliot Quataert, Martin Rees, Joop Schaye, Joe Silk, Francesco Tombesi, and 
Sylvain Veilleux.


\end{document}